\documentclass[twocolumn,trackchanges]{aastex62}

\usepackage{amsmath}
\usepackage{xspace}

\newcommand{\unitspace}{\ensuremath{\mskip\thinmuskip}}
\newcommand{\msun}{\ensuremath{M_{\odot}}\xspace}
\newcommand{\gpercc}{\ensuremath{{\mathrm{g}\unitspace\mathrm{cm}^{-3}}}}
\newcommand{\degree}{\ensuremath{^{\circ}}\xspace}

\newcommand{\nue}{\ensuremath{\nu_{e}}\xspace}
\newcommand{\nubar}{\ensuremath{\bar{\nu}_{e}}\xspace}
\newcommand{\nux}{\ensuremath{\nu_{x}}\xspace}

\newcommand{\dint}{\ensuremath{\unitspace\mathrm{d}}}

\received{-}
\revised{-}
\accepted{-}
\submitjournal{ApJ}
\shorttitle{3D Supernova Simulations: FMD vs. RbR+}
\shortauthors{Glas et al.}

\begin{document}

\title{Three-Dimensional Core-Collapse Supernova Simulations with Multi-Dimensional Neutrino Transport\\
Compared to the Ray-by-Ray-plus Approximation}

\correspondingauthor{Robert Glas}
\email{rglas@mpa-garching.mpg.de}

\author[0000-0002-7040-9472]{Robert~Glas}
\affiliation{Max-Planck-Institut f\"ur Astrophysik, Karl-Schwarzschild-Stra{\ss}e 1, D-85748 Garching, Germany}
\affiliation{Physik Department, Technische Universit\"at M\"unchen, James-Franck-Stra{\ss}e 1, D-85748 Garching, Germany}

\author[0000-0002-3126-9913]{Oliver~Just}
\affiliation{Astrophysical Big Bang Laboratory, RIKEN Cluster for Pioneering Research, 2-1 Hirosawa, Wako, Saitama 351-0198, Japan}
\affiliation{Max-Planck-Institut f\"ur Astrophysik, Karl-Schwarzschild-Stra{\ss}e 1, D-85748 Garching, Germany}

\author[0000-0002-0831-3330]{H.-Thomas~Janka}
\affiliation{Max-Planck-Institut f\"ur Astrophysik, Karl-Schwarzschild-Stra{\ss}e 1, D-85748 Garching, Germany}

\author[0000-0001-5664-1382]{Martin~Obergaulinger}
\affiliation{Institut f\"ur Kernphysik, Technische Universit\"at Darmstadt, Schlossgartenstra{\ss}e 2, 64289 Darmstadt, Germany.}
\affiliation{Departament d'Astronomia i Astrof{\'i}sica, Universitat de Val{\`e}ncia, Edifici d{\'{}}Investigaci{\'o} Jeroni Mu{\~n}oz, \\
C/ Dr.~Moliner, 50, E-46100 Burjassot (Val{\`e}ncia), Spain}

\begin{abstract}

Self-consistent, time-dependent supernova (SN) simulations in three spatial dimensions
(3D) are conducted with the \textsc{Aenus-Alcar} code, comparing, for the first time,
calculations with fully multi-dimensional (FMD) neutrino transport and the ray-by-ray-plus
(RbR+) approximation, both based on a two-moment solver with algebraic M1 closure.
We find good agreement between 3D results with FMD and RbR+ transport for both tested
grid resolutions in the cases of a 20\,$M_\odot$ progenitor, which does not explode with
the employed simplified set of neutrino opacities, and of an exploding 9\,$M_\odot$ model.
This is in stark contrast to corresponding axisymmetric (2D) simulations, which confirm
previous claims that the RbR+ approximation can foster explosions in 2D in particular in models
with powerful axial sloshing of the stalled shock due to the standing accretion shock
instability (SASI). However, while local and instantaneous variations of neutrino fluxes
and heating rates can still be considerably higher with RbR+ transport in 3D, the
time-averaged quantities are very similar to FMD results because of the absence of a
fixed, artificial symmetry axis that channels the flow. Therefore, except for stochastic
fluctuations, the neutrino signals and the post-bounce evolution of 3D simulations with
FMD and RbR+ transport are also very similar, in particular for our calculations with the
better grid resolution. Higher spatial resolution has clearly a more important impact
than the differences by the two transport treatments. Our results back up the use of the
RbR+ approximation for neutrino transport in 3D SN modeling.

\end{abstract}


\section{Introduction}\label{sec:introduction}

Stars with masses larger than ${\sim 8\unitspace\msun}$ collapse
at the end of their lives, leaving behind
either a neutron star or a black hole.
The death of these stars is often accompanied
by a core-collapse supernova (CCSN)
that is capable of outshining an entire galaxy in visible light.
However, the mechanism driving these explosions is still a matter of active debate.
The initially launched shock wave has been shown to quickly lose energy
and to stall inside the iron core of the progenitor star.
According to the delayed neutrino-driven mechanism,
neutrinos streaming out of the cooling proto-neutron star
help to revive the shock
by depositing energy and triggering hydrodynamical instabilities
in the layer behind the shock wave
\citep[for extensive reviews, see][]{2005ARNPS..55..467M,2012AdAst2012E..39K,2012ARNPS..62..407J,2015PASA...32....9F,
2016ARNPS..66..341J,2016PASA...33...48M}.

Detailed studies of the CCSN mechanism rely
on first-principle simulations of the post-bounce phase
that self-consistently describe the evolution of the stellar fluid
together with the transport of neutrinos,
including self-gravity and a microphysical equation of state
\citep[see, e.g.][]{2016ARNPS..66..341J,2016PASA...33...48M}.
Simulations in spherical symmetry (one-dimensional, 1D) failed to explain
core-collapse supernovae for most cases
\citep[e.g.][]{2003ApJ...592..434T},
but the neutrino-heating mechanism has been shown to work in many axisymmetric (two-dimensional, 2D)
and a growing suite of three-dimensional (3D) simulations
\citep{2012PTEP.2012aA309J,2012ApJ...749...98T,2014ApJ...786...83T,2015ApJ...807L..31L,2015MNRAS.453..287M,2016ARNPS..66..341J,
2016ApJ...818..123B,2016ApJ...831...98R,2016ApJ...825....6S,2018ApJ...855L...3O}.
However, explosions in 2D are artificially influenced by the imposed symmetry axis,
and 3D simulations do not yet show robust explosions
\citep[e.g.][]{2013ApJ...770...66H,2015ApJ...808L..42M,2017MNRAS.472..491M,2018ApJ...852...28S,2018ApJ...865...81O},
except with significantly simplified neutrino-transport physics \citep[e.g.][]{2018ApJ...855L...3O}
or with artificially imposed progenitor perturbations \citep{2019MNRAS.482..351V}.

One important aspect of CCSN modeling is the transport of neutrinos,
which is governed by the seven-dimensional (3 spatial and 3 momentum dimensions plus time) Boltzmann transport equation
describing the time evolution of the single-particle distribution
function in the six-dimensional phase space.
Solving it in its full complexity, for example by the discrete-ordinate ($S_{n}$) method \citep{2012ApJS..199...17S},
is currently computationally only feasible for selected temporal snapshots \citep{2015ApJS..216....5S},
or for time-dependent axisymmetric hydrodynamic simulations with low resolution in momentum space \citep{2018ApJ...854..136N}.
Alternatively, a Monte Carlo solver (Sedonu) has been applied for comparison with the $S_{n}$ solutions on a 2D stationary background
\citep{2017ApJ...847..133R}.

For this reason, a variety of approximations have been introduced to reduce the complexity of the full Boltzmann equation.
The isotropic diffusion-source approximation is based on the decomposition of the neutrino distribution function into a trapped particle
and a streaming particle component, which are separately evolved \citep{2009ApJ...698.1174L,2016ApJ...817...72P}.
Other schemes evolve the angular moments (i.e., integrals) of the specific intensity of neutrinos,
instead of the particle phase-space distribution function itself.
In flux-limited diffusion schemes, for example, the neutrino energy density (corresponding to the zeroth angular moment) is evolved,
and the neutrino flux density (first moment) is obtained from the gradient of the energy density
\citep{1985ApJS...58..771B,2013ApJ...767L...6B,2013ApJS..204....7Z}.
Two-moment transport schemes, which also evolve the neutrino flux density,
need to close the system of moment equations by relating the Eddington tensor (second moment) to the low-order moments
\citep[often called ``M1'' methods and used in many published codes;][]
{2015PhRvD..91l4021F,2015MNRAS.453.3386J,2018ApJ...854...63O,2016ApJ...831...81S,2016ApJ...831...98R,2016ApJS..222...20K}.
Another widely-used approximation is ray-by-ray-plus (RbR+) transport,
which assumes the specific intensity to be symmetric around the radial direction
of the spherical polar coordinate grid \citep{2002A&A...396..361R,2006A&A...447.1049B,2013ApJ...767L...6B}.
A direct consequence of RbR+ are vanishing lateral and azimuthal (i.e., non-radial) flux components.
All of these approximations require tests of their viability by mutual comparison and, ultimately, comparison with full Boltzmann solutions.

There have been speculations about the influence of RbR+ on the neutrino-heating mechanism in 2D,
mostly based on studies that neglect time-dependent effects, leaving the temporal evolution unclear.
\citet{2015ApJ...800...10D} analyzed neutrino fluxes based on a temporal snapshot of a two-dimensional CCSN simulation.
They found larger angular variations of the radiation fluxes for RbR+
compared to their multi-group, flux-limited diffusion scheme in 2D,
and they speculated that RbR+ might exaggerate angular variations
because of an unphysically high correlation between matter and the neutrino field.
Another study by \citet{2015ApJS..216....5S} investigated neutrino-heating rates
in the gain layer (i.e. where neutrino-matter interactions result in net heating),
based on temporal snapshots of 3D CCSN simulations.
When comparing results from solving the Boltzmann equation directly to results with the RbR+ approximation,
they found differences in the local heating rates of 20\%.
However, the rates integrated spatially over the gain layer differed by only 2\%
\citep[a similar conclusion was reached by][]{2006A&A...447.1049B}.
A time-dependent investigation was carried out by \citet{2016ApJ...831...81S},
who performed self-consistent, two-dimensional CCSN simulations of the post-bounce phase of various progenitor models.
They hypothesize that there may be an unphysically strong feedback
between the axial sloshing of the standing accretion shock instability \citep[SASI,][]{2003ApJ...584..971B} and the neutrino field,
leading to explosions in axisymmetric simulations with RbR+ more readily,
compared to applications employing a multi-group, fully multi-dimensional two-moment transport solver.
In addition, \citet{2018MNRAS.481.4786J} found the RbR+ approximation to be conducive to explosions in time-dependent 2D CCSN simulations
in particular for models that exhibit violent activity of SASI,
when compared to a fully multi-dimensional two-moment transport scheme.
The 2D behavior, however, turned out to be highly stochastic and very sensitive to minor variations of the applied input physics,
for which reason the 2D results were hard to interpret in a clear way.

So far, time-dependent studies have investigated the influence of RbR+ only in 2D simulations,
in which hydrodynamical instabilities are constrained by the imposed symmetry axis.
Furthermore, 3D simulations allow for different modes of instabilities \citep[e.g. the spiral mode of the SASI,][]{2007ApJ...656..366B},
and they show a different transport of turbulent energy compared to the 2D case \citep{2012ApJ...755..138H,2013ApJ...775...35C,2015ApJ...799....5C}.

For these reasons, it is indispensable to assess the influence of the neutrino-transport treatment on the heating mechanism
also in time-dependent 3D simulations.
We fulfill this need by comparing self-consistent, time-dependent CCSN simulations in three dimensions
with a fully multi-dimensional (FMD) neutrino-transport scheme and with the RbR+ approximation.

This paper is structured as follows:
We describe the numerical method and the setup of our simulations in Section~\ref{sec:simulation_setup},
followed by a short summary of two-dimensional simulations in Section~\ref{sec:two-dimensional}.
In Section~\ref{sec:three-dimensional} we present our main findings for 3D simulations with FMD and RbR+ neutrino transport in comparison.
We conclude in Section~\ref{sec:conclusion}.

\section{Simulation Setup}\label{sec:numerics}\label{sec:simulation_setup}\label{sec:setup}

All CCSN simulations presented in this paper were conducted with the radiation-hydrodynamics code
\textsc{Aenus-Alcar} \citep{2008ObergaulingerPhD,2015MNRAS.453.3386J,2018MNRAS.481.4786J}.
It solves the hydrodynamics equations \citep[i.e., Equations ({1a-d}) in][]{2018MNRAS.481.4786J}
which describe the conservation of mass, momentum, total (internal plus kinetic) energy,
and electron fraction of the stellar fluid.
The hydrodynamics solver is based on a Godunov-type, directionally unsplit finite-volume scheme
in spherical polar coordinates (with radius $r$, polar angle $\theta$, and azimuth angle $\phi$)
and employs high-order reconstruction methods with an approximate Riemann solver to obtain cell-interface fluxes.
The time integration is realized by an explicit, second-order Runge-Kutta scheme with a time step
that is constrained by the Courant-Friedrichs-Lewy condition \citep[CFL,][]{1928MatAn.100...32C}.
The hydrodynamics equations are closed by the microphysical equation of state (SFHo) from \citet[][]{2013ApJ...774...17S},
which we extended to a minimum temperature of ${10^{-3}\unitspace\mathrm{MeV}}$
to describe low-temperature regions that we encounter in a low-mass progenitor model.
The code accounts for the gravitational self-interaction of the fluid by solving Poisson's equation
(including general relativistic corrections, see \citealt{2006A&A...445..273M}, case A) in spherical symmetry.

The hydrodynamics module is coupled to an energy-dependent radiation-transport solver for neutrinos,
which is based on an FMD two-moment scheme that evolves the zeroth and first angular moments of the Boltzmann transport equation
to solve for the energy density ${E_{\nu}}$ and energy-flux density ${F^{i}_{\nu}}$ (with ${i \in \{r,\theta,\phi\}}$) of neutrinos $\nu$
\citep[i.e., Equations ({3a-b}) in][]{2018MNRAS.481.4786J}.
Both ${E_{\nu}}$ and ${F^{i}_{\nu}}$ are measured in the co-moving frame of the stellar fluid.
We evolve three neutrino species (electron neutrinos \nue, anti-electron neutrinos \nubar,
and a third species \nux that represents all four heavy-lepton neutrinos)
and for the neutrino energy $\varepsilon$ we employ 15 bins
that are logarithmically spaced in the interval ${0 \le \varepsilon \le 400\unitspace\mathrm{MeV}}$.
The evolved equations are closed by computing the higher moments (e.g.\ the radiation pressure tensor)
via an algebraic relation that depends on the zeroth and first moments.
We include all velocity-dependent terms (e.g.\ the Doppler shift terms),
which appear in the co-moving frame representation of the moment equations, to an accuracy of order $v/c$.
In analogy to the hydrodynamics module, we use a Godunov-type, finite-volume scheme to solve the system of hyperbolic moment equations.
The neutrino-matter interactions are implemented via source terms, which couple the neutrino moment equations to the hydrodynamics equations
\citep[i.e., Equations ({2a-c}) in][]{2018MNRAS.481.4786J}.
In particular, contributions of neutrino interactions enter the equations for energy (${Q_{\mathrm{E}}}$),
momentum (${Q_{\mathrm{M}}}$), and electron fraction (${Q_{\mathrm{N}}}$) of the stellar matter.
The integration of the moment equations is performed explicitly in time (with the same time stepping as for the hydrodynamics equations)
for most terms except for some interaction rates, whose source terms are treated implicitly in time
\citep[for detailed explanations, see appendix A of][]{2018MNRAS.481.4786J}.

\textsc{Alcar}'s set of neutrino-matter interactions contains absorption and emission of neutrinos by nucleons (neutrons and protons)
for electron-type neutrinos (\nue and \nubar),
and isoenergetic scattering on nucleons for all neutrino species \citep{1985ApJS...58..771B,1993ApJ...405..637M}.
All interaction cross sections with nucleons include weak-magnetism and nucleon-recoil corrections \citep{2002PhRvD..65d3001H}.
We further take into account absorption and emission of \nue by nuclei \citep{1985ApJS...58..771B,1993ApJ...405..637M},
and coherent scattering on nuclei with ion-screening corrections due to medium correlations for all three species
\citep{1997PhRvD..56.7529B,1997PhRvD..55.4577H}.
We also include inelastic scattering of neutrinos off electrons and positrons \citep{1977ApJ...217..565Y,1985ApJS...58..771B,1994ApJ...433..247C}
and follow a suggestion by \citet{2015ApJS..219...24O} in damping the scattering source terms
at densities larger than ${5\times 10^{12}\unitspace\gpercc}$
when the scattering rates become too high to be followed properly with an explicit time-stepping scheme.
The rates for nucleon-nucleon bremsstrahlung \citep{1998ApJ...507..339H}
and neutrino-antineutrino pair-production by $e^{+}e^{-}$ annihilation \citep{1985ApJS...58..771B,1998A&AS..129..343P}
are implemented for \nux with a simplified description,
in which these interactions are treated similarly to absorption and emission terms \citep[for details see][]{2015ApJS..219...24O}.

To perform comparisons between the FMD neutrino-transport scheme and the RbR+ approximation,
we set the non-radial neutrino-flux components (${F^{\theta}_{\nu}}$ and ${F^{\phi}_{\nu}}$)
to zero for simulations that employ the RbR+ approximation,
retaining only the evolution of the radial flux component (${F^{r}_{\nu}}$).
In all of our simulations, we use the advanced ray-by-ray-plus version \citep[][]{2006A&A...447.1049B},
which includes both the non-radial velocity-dependent terms in the neutrino moment equations and,
at densities above ${10^{12}\unitspace\gpercc}$,
the non-radial neutrino-pressure contributions in the hydrodynamics equations.

To assess the influence of numerical grid resolution,
we vary the number of grid cells ($N_{r}$, $N_{\theta}$, and $N_{\phi}$, respectively).
A summary of our 2D and 3D simulations can be found in Table~\ref{tab:grids}.
In general, we differentiate between ``low'' and ``high'' resolutions.
For 2D simulations our naming convention is based on the number of lateral zones
(i.e., ``40'' for low resolution, and ``80'' for high resolution),
whereas for 3D simulations we simply abbreviate low (``L'') and high (``H'') resolutions.
In both 2D and 3D simulations,
the radial grid ranges from 0 to ${10^{9}\unitspace\textrm{cm}}$ and contains ${N_{r} \in \left\{ 320, 640 \right\}}$ zones.
The cell sizes in radial direction increase logarithmically, starting with ${10^{4}\unitspace\textrm{cm}}$ in the center of the star.
In radial direction, we use a reflecting inner boundary (at ${r = 0}$) for all evolved quantities,
and at the outer boundary (at ${r = 10^{9}\unitspace\textrm{cm}}$) we use an inflow condition for hydrodynamic quantities
with time-independent values for density, infall velocity, and total energy (all obtained from the progenitor model)
and prescribe a free outflow condition for the neutrinos.

\begin{deluxetable}{cccc}
\tablecaption{Summary of simulation parameters.\label{tab:grids}}
\tablewidth{0pt}
\tablehead{
    \colhead{} & \multicolumn{1}{c}{Label} & \colhead{Angular Grid Structure\tablenotemark{a}} &
    \colhead{$N_{r} \times N_{\theta}$ ($\times N_{\phi}$)}
}
\startdata
3D    & L        & polar      & $320 \times 40 \times 80\phantom{0}$  \\
3D    & H        & polar      & $640 \times 80 \times 160$ \\ \hline
2D    & pol/40   & polar      & $320 \times 40$            \\
2D    & pol/80   & polar      & $640 \times 80$            \\
2D    & equ/80   & equatorial & $640 \times 80$            \\
2D    & uni/80   & uniform    & $640 \times 80$            \\
\enddata
\tablenotetext{a}{We distinguish between ``polar'' (coarser resolution at the poles),
``equatorial'' (coarser resolution at the equator),
and ``uniform'' (uniform resolution) angular grid structures.}
\end{deluxetable}

In 3D simulations, the polar grid is composed of two large zones at each pole (${\theta = 0}$ and ${\theta = 180^{\circ}}$, respectively)
and ${N_{\theta} - 4}$ zones around the equatorial plane with equal sizes.
For low-resolution grids with ${N_{\theta} = 40}$ zones (high with ${N_{\theta} = 80}$)
the first cell measures ${12^{\circ}}$ (${10^{\circ}}$) and the second cell ${6^{\circ}}$ (${4^{\circ}}$),
leading to a cell size of ${4^{\circ}}$ (${2^{\circ}}$) for the remaining equally sized zones to cover the entire ${180^{\circ}}$ range.
Using this particular arrangement with coarser angular resolution at the poles, we are able to sustain a computationally affordable time step
by avoiding exceedingly small cell sizes in azimuthal direction, which are proportional to ${\mathrm{sin}\unitspace\theta}$,
and which dictate the time step in our 3D simulations according to the CFL criterion.
In azimuthal direction (relevant only in 3D), the grid cells have constant size over the whole ${360^{\circ}}$ domain,
leading to a cell size of ${4.5^{\circ}}$ (${2.25^{\circ}}$) for ${N_{\phi} = 80}$ (${N_{\phi} = 160}$) zones.
We use a reflecting boundary in lateral direction (at angles ${\theta = 0\degree}$ and ${\theta = 180\degree}$) for 2D and 3D simulations,
and periodic boundary conditions in azimuthal direction (at the interface where ${\phi = 0\degree}$ or, equivalently, ${\phi = 360\degree}$)
for 3D simulations.

In our 2D simulations, we assess the influence of the polar grid by varying the number and arrangement of angular grid cells.
First, we performed 2D simulations on a polar grid with ${N_{\theta} = 80}$ uniformly distributed cells
(labeled `uni/80', see Table~\ref{tab:grids}), leading to a polar grid resolution of ${2.25^{\circ}}$.
Second, we tested the polar grid structure that we apply in the 3D simulations
(with coarser angular resolution at the poles to alleviate time step constraints)
also in 2D simulations (labeled `pol/40' for ${N_{\theta} = 40}$, and `pol/80' for ${N_{\theta} = 80}$ zones, respectively).
Third, to further assess the influence of partially coarsened angular resolution,
we conducted one 2D simulation on a polar grid with coarse resolution at the equator (labeled `equ/80' with ${N_{\theta} = 80}$ zones).
Specifically, the `equ/80' grid consists of two ${10^{\circ}}$-zones symmetrically arranged around the equator (${\theta = 90^{\circ}}$),
surrounded by two ${4^{\circ}}$-zones and 76 remaining zones with ${2^{\circ}}$ to cover the entire ${180^{\circ}}$ range.

In both 2D and 3D simulations, further constraints on the time step arise near the center of the spherical coordinate system,
where cell sizes in lateral and azimuthal directions are smallest because of their dependence on the radial coordinate.
We alleviate these constraints by evolving the innermost ${10^{6}\unitspace\mathrm{cm}}$ in spherical symmetry and, thus,
neglecting angular variations in the (spherically symmetric) innermost region of the proto-neutron star.
This is an acceptable approach as long as the central core of ${r < 10^{6}\unitspace\mathrm{cm}}$ remains convectively stable,
which is practically fulfilled until about ${100\unitspace\mathrm{ms}}$ after bounce
but can become marginally valid afterwards.
While the violation of this condition has some influence on the evolution of proto-neutron star convection,
the consequences are irrelevant for the results discussed in the present paper.

Because there is no growth of hydrodynamic instabilities during core collapse until bounce,
we simulate the collapse of the iron core in 1D and start our 2D and 3D simulations
from the 1D data at ${15\unitspace\mathrm{ms}}$ after core bounce.
The 1D simulations are performed on the same radial grid (with ${N_{r} = 640}$) that we use in 2D and 3D simulations.
To trigger hydrodynamic instabilities in the post-bounce phase, we initialize the multi-dimensional simulations
with random seed perturbations in the density with an amplitude of 0.1\%.

\begin{figure}
    \includegraphics[width=0.48\textwidth]{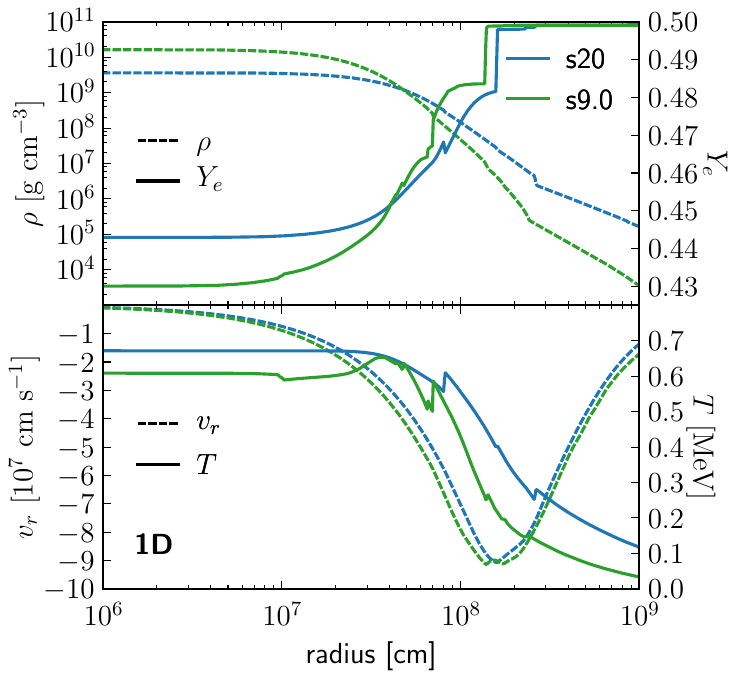}
    \caption{Radial profiles at the onset of collapse of the spherically symmetric progenitor models s20 (blue lines) and s9.0 (green lines).
            We show, in dependence on the radius $r$, the densities $\rho$ and electron fractions $Y_{e}$ in the top panel,
            and the radial velocities $v_{r}$ and temperatures $T$ in the bottom panel.
            \label{fig:comparison_1D_progenitors}}
\end{figure}

We conducted CCSN simulations for two different non-rotating, solar-metallicity progenitor models
(see Figure~\ref{fig:comparison_1D_progenitors} for the radial profiles of
density, electron fraction, radial velocity, and temperature at the onset of collapse).
The first model, a ${20\unitspace\msun}$ star from \citet[][labeled s20 in the following]{2007PhR...442..269W},
has already been extensively studied in literature.
In 2D, both successful \citep{2016ApJ...818..123B,2016ApJ...825....6S}
and failed explosions \citep{2016ApJ...831...81S,2018MNRAS.477.3091V} have been reported.
Other groups have found successful and failed explosions in dependence on the simulation setup
(e.g., the employed set of neutrino-interaction rates)
when performing 2D \citep[see, e.g.,][]{2017PhRvL.119x2702B,2018ApJ...853..170K,2018ApJ...854...63O,2018MNRAS.481.4786J}
and 3D \citep{2015ApJ...808L..42M} simulations of the s20 progenitor model.
Thus, the s20 progenitor model seems to reside at the threshold between successful and failed explosions.

\begin{figure}
    \includegraphics[width=0.48\textwidth]{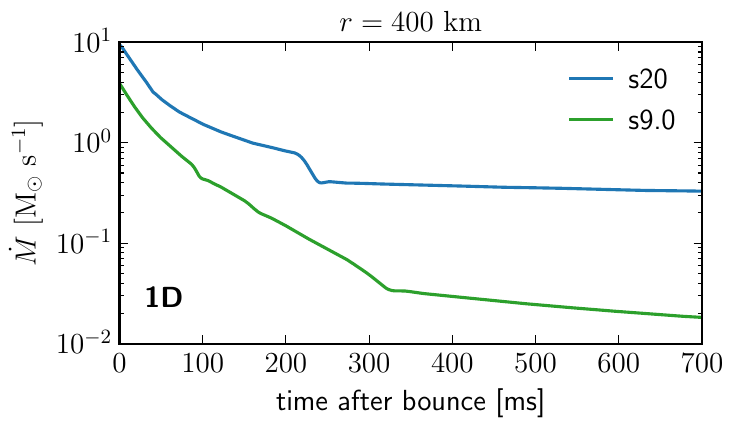}
    \caption{Mass accretion rates as functions of the time after bounce for 1D simulations
            of the progenitor models s20 (blue line) and s9.0 (green line).
            The mass accretion rates $\dot{M}$ (see Equation~\eqref{eqn:mass_accretion_rate}) are
            evaluated at a radius of $r=400\unitspace\mathrm{km}$.
            Note that $\dot{M}$ shows excellent agreement between 1D, 2D, and 3D simulations
            (until the time of shock revival in the case of successful explosions).
            \label{fig:comparison_1D_progenitors_mdot}}
\end{figure}

In order to additionally compare the explosion dynamics of 2D and 3D simulations,
we chose the second progenitor model such that we expect robust successful explosions.
The ${9\unitspace\msun}$ progenitor model from \citet[][labeled s9.0 in the following]{2015ApJ...810...34W}\footnote{We
use a slightly modified version of the ${9\unitspace\msun}$ progenitor model,
which is described by \citet{2016ApJ...821...38S}.
Data are available at
\url{https://wwwmpa.mpa-garching.mpg.de/ccsnarchive/data/SEWBJ_2015/index.html}.}
has been shown to rather readily explode in 2D simulations by \citet{2017ApJ...850...43R,2018MNRAS.481.4786J}.
As most models at the low-mass end of iron-core progenitors,
the s9.0 star exhibits a very steep density gradient at the edge of the degenerate core
(see the radial density profile in Figure~\ref{fig:comparison_1D_progenitors}).
As a consequence of the steep density gradient,
the mass accretion rate drops rapidly during the first few ${100\unitspace\mathrm{ms}}$ after core bounce,
leading to favorable conditions for shock runaway.
This can be seen from Figure~\ref{fig:comparison_1D_progenitors_mdot},
which shows the mass accretion rate, evaluated at a radius of ${r = 400\unitspace\mathrm{km}}$,
for the 1D simulations of the s9.0 and s20 progenitor models.
Here, the mass accretion rate at any radius $r$ is defined by the surface integral,
\begin{equation}\label{eqn:mass_accretion_rate}
\dot{M} (r) =
r^{2} \int_{4\pi} \rho \; v_{r} \dint{\Omega},
\end{equation}
with baryonic mass density $\rho$ and radial velocity component $v_{r}$.

\section{Two-Dimensional Simulations}\label{sec:two-dimensional}
In this section we summarize the results of our 2D simulations for the s20 and s9.0 progenitor models.
We will make use of these results in Section~\ref{sec:three-dimensional}, where we will compare 2D and 3D simulations.
Since this paper mainly focuses on 3D results, we refer the reader to \citet{2018MNRAS.481.4786J},
who performed a detailed comparison of 1D and 2D CCSN simulations using the \textsc{Alcar} code
and varying the physics inputs in a large set of models.
Their results regarding the comparison between FMD and RbR+ neutrino transport in 2D simulations are consistent
with our 2D results summarized in this section.

Here, we present only a small sample of 2D simulations with either FMD neutrino transport or the RbR+ approximation.
To assess the influence of grid resolution, we varied the number of radial and angular grid zones
(the labels of our simulations contain the number of angular grid zones, see Section~\ref{sec:numerics} and Table~\ref{tab:grids}).
For the s20 progenitor model, we also varied the type of the polar grid.
Besides uniform angular cell spacing (labeled `uni'),
coarser resolution at the poles (labeled `pol' and
copying the lateral grid structure applied to 3D simulations to alleviate time step constraints) was tested.
Moreover, we performed one simulation on a polar grid with coarser resolution at the equator (labeled `equ').
Except for using different neutrino-transport methods and the aforementioned grid parameters,
all physical and numerical inputs are identical in our simulations.

\begin{figure}
    \includegraphics[width=0.48\textwidth]{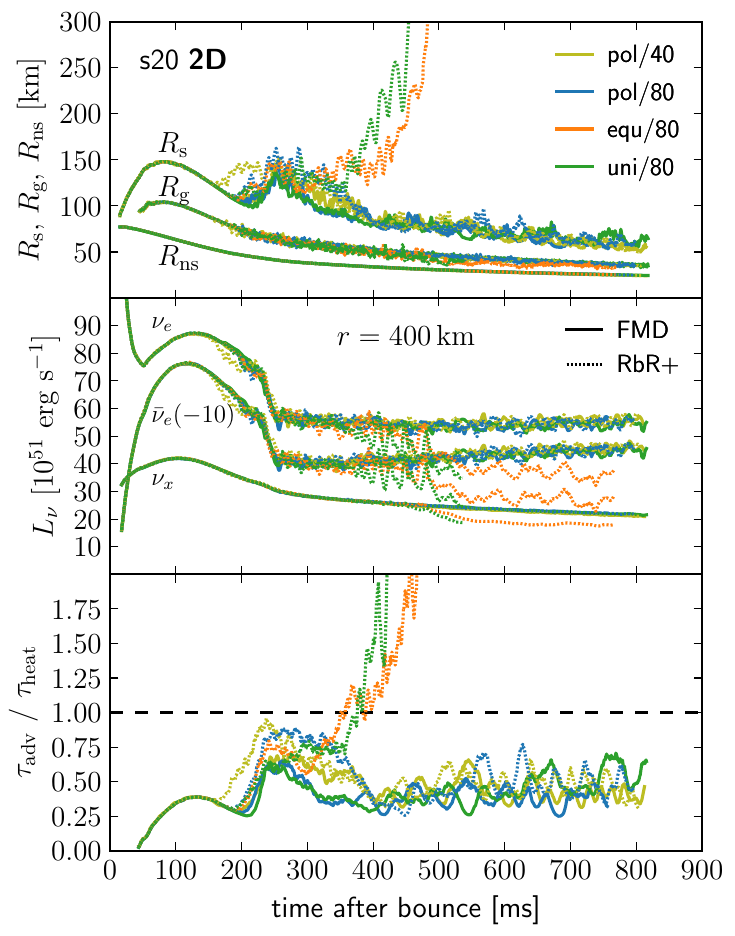}
    \caption{Comparison of 2D simulations of the s20 progenitor model.
        We show, as functions of the time after bounce, the angle-averaged shock radii $R_{\mathrm{s}}$,
        gain radii $R_{\mathrm{g}}$, and neutron-star radii $R_{\mathrm{ns}}$ (all in the top panel; see labels),
        the neutrino luminosities $L_{\nu}$ (central panel) for \nue, \nubar, and \nux (see labels),
        and the ratios of advection to heating timescales $\tau_{\mathrm{adv}} / \tau_{\mathrm{heat}}$ (bottom panel).
        The neutrino luminosities are evaluated at a radius of ${r = 400\unitspace\mathrm{km}}$ in the co-moving frame of the stellar fluid.
        The simulations differ in the neutrino-transport scheme (line style),
        and in the grid type and resolution (line color), see legend and Table~\ref{tab:grids} for an overview.
        Note that the neutrino luminosities and the ratios of advection to heating timescales
        were smoothed by running averages of ${5\unitspace\mathrm{ms}}$,
        and that some lines were shifted vertically to facilitate readability of the plot.
        These lines are marked with labels indicating the number by which they are shifted in units of the ordinate.
        \label{fig:s20_comparison_2D}}
\end{figure}

We present an overview of all 2D simulations for the s20 progenitor model in Figure~\ref{fig:s20_comparison_2D}.
The top panel shows the temporal evolution of the angle-averaged shock radii ${R_{\mathrm{s}}}$,
which indicate successful shock revival for two of the RbR+ simulations
between ${400\unitspace\mathrm{ms}}$ and ${500\unitspace\mathrm{ms}}$ after core bounce.
In these simulations, shock revival occurs shortly after the Si/Si-O interface falls through the shock,
leading to a significant drop in the mass accretion rate at ${\sim200\unitspace\mathrm{ms}}$
(compare Figure~\ref{fig:comparison_1D_progenitors_mdot}).
As the central panel of Figure~\ref{fig:s20_comparison_2D} shows,
successful shock revival is accompanied by a drop in the neutrino luminosities,
which is a direct consequence of the reduced mass accretion onto the neutron star.
We define the luminosity $L_{\nu}$ for any neutrino species $\nu$ at a radius $r$ in the co-moving frame of the stellar fluid by
\begin{equation}\label{eqn:neutrino_luminosity}
L_{\nu} (r) = r^{2} \int F^{r}_{\nu} (r) \dint{\varepsilon} \dint{\Omega} .
\end{equation}
In contrast, non-exploding models exhibit almost constant neutrino luminosities after ${200\unitspace\mathrm{ms}}$ post bounce (p.b.)
because of continuing mass accretion.
From the evolution of the shock radii we also conclude that none of the 2D simulations with FMD neutrino transport
results in an explosion until at least ${700\unitspace\mathrm{ms}}$ after bounce.

As already argued by \citet{2015ApJ...800...10D} and \citet{2016ApJ...831...81S},
this discrepancy between 2D simulations with RbR+ and FMD neutrino transport may be explained by
a stronger feedback between the neutrino field and the axial sloshing of the SASI in the case of the RbR+ approximation.
For this reason, we analyzed the time-dependent neutrino-heating rates at the poles,
and we found deviations from the angle-averaged values to be significantly larger in RbR+ simulations.
We will come back to a discussion of this behavior in our comparison between differences of 2D and 3D heating rates (see Section~\ref{sec:s20_3D}).
This analysis confirms that in 2D simulations with strong SASI sloshing along the axis,
the RbR+ approximation can amplify local variations at the poles and,
therefore, may lead to conditions that are more beneficial for shock revival
than corresponding results with an FMD neutrino-transport scheme.

In addition to discrepancies between 2D simulations with RbR+ and FMD neutrino transport,
we find further differences in dependence on the polar grid.
As the top panel of Figure~\ref{fig:s20_comparison_2D} shows,
we find successful shock revival only for RbR+ simulations with high angular grid resolution at the poles
(labels `uni/80' and `equ/80').
In contrast, simulations that were performed on a grid with coarse polar resolution (labels `pol/40' and `pol/80') do not exhibit shock revival.

These discrepancies do not seem to be caused by differences in the time-dependent evolution
of the angle-integrated neutrino quantities.
Until the time when exploding models experience shock runaway,
we find the neutrino luminosities, mean energies, and root mean square energies for all neutrino species
to be in good agreement between all 2D simulations.
The central panel of Figure~\ref{fig:s20_comparison_2D} shows only, for an example, the temporal evolution of the neutrino luminosities
for all neutrino species.

To assess the influence of the angular grid resolution, we analyze the conditions for shock revival in our 2D simulations.
Favorable conditions for an explosion arise when matter remains in the gain layer for a time that is long enough
for neutrinos to deposit sufficient energy to cause shock expansion.
The total binding energy of matter in the gain layer, ${E^{\mathrm{g}}_{\mathrm{tot}}}$,
is calculated from the specific internal energy $e_{\mathrm{int}}$,
specific kinetic energy ${v^{2} / 2}$, and gravitational potential $\phi_{\mathrm{grav}}$ as
\begin{equation}\label{eqn:total_energy_gain}
E^{\mathrm{g}}_{\mathrm{tot}} =
\int_{V_{\mathrm{gain}}} \rho
\left( e_{\mathrm{int}} + \frac{v^{2}}{2} + \phi_{\mathrm{grav}} \right) \dint V ,
\end{equation}
with velocity magnitude ${v}$.
The volume of the gain layer, ${V_{\mathrm{gain}}}$, encompasses the region from the average gain radius ${R_{\mathrm{g}}}$,
at which the angle-averaged net neutrino energy transfer turns from cooling into heating,
to the angle-dependent shock radius ${R_{\mathrm{s}} (\theta, \phi)}$.
The characteristic timescale for neutrino heating is then given by
\citep{2006A&A...457..281B,2012ApJ...756...84M,2018ApJ...852...28S}
\begin{equation}\label{eqn:heating_timescale}
\tau_{\mathrm{heat}} = \frac{|E^{\mathrm{g}}_{\mathrm{tot}}|}{Q^{\mathrm{g}}_{\nu}} ,
\end{equation}
with ${Q^{\mathrm{g}}_{\nu}}$ being the total neutrino-heating rate in the gain layer,
\begin{equation}\label{eqn:heating_gain_layer}
Q^{\mathrm{g}}_{\nu} =
\int_{V_{\mathrm{gain}}} Q_{E} \dint V .
\end{equation}
Here, ${Q_{E}}$ is the energy source term (energy exchange rate per volume)
which couples the hydrodynamics and neutrino-transport equations (compare \citealt{2018MNRAS.481.4786J}, their Equation (2c)).
The typical timescale for matter residing in the gain layer is given by the advection timescale,
which we approximate \citep[assuming steady-state conditions,][]{2009ApJ...694..664M,2012ApJ...756...84M} by
\begin{equation}\label{eqn:advection_timescale}
\tau_{\mathrm{adv}} \approx
\frac{M_{\mathrm{g}}}{\dot{M}} ,
\end{equation}
with $M_{\mathrm{g}}$ being the total mass in the gain layer,
\begin{equation}\label{eqn:mass_gain_layer}
M_{\mathrm{g}} =
\int_{V_{\mathrm{gain}}} \rho \dint V ,
\end{equation}
and ${\dot{M}}$ being the mass accretion rate (see Equation~\eqref{eqn:mass_accretion_rate}),
evaluated at a radius of ${r = 400\unitspace\mathrm{km}}$.
In the case the ratio of advection to heating timescale, ${\tau_{\mathrm{adv}} / \tau_{\mathrm{heat}}}$, exceeds unity,
neutrino heating leads to beneficial conditions for shock expansion and may ultimately power an explosion
\citep{2001A&A...368..527J,2005ApJ...620..861T,2012ApJ...749..142F}.

We show the ratios ${\tau_{\mathrm{adv}} / \tau_{\mathrm{heat}}}$ in the bottom panel of Figure~\ref{fig:s20_comparison_2D}.
At ${400\unitspace\mathrm{ms}}$ p.b., approximately when shock expansion sets in,
we find exploding models to reach and exceed ratios of unity.
Non-exploding simulations with RbR+ (`pol/40' and `pol/80') exhibit slightly larger timescale ratios than the exploding models
in the phase between ${200\unitspace\mathrm{ms}}$ and ${300\unitspace\mathrm{ms}}$ p.b.,
reaching values close to unity in this phase without, however, leading to successful shock revival.
This strongly suggests that the explosions are caused by pole effects in models with fine resolution at the poles.
In contrast, the values for ${\tau_{\mathrm{adv}} / \tau_{\mathrm{heat}}}$ stay well below unity
in all (non-exploding) simulations with FMD neutrino transport.
This confirms that RbR+ simulations exhibit more favorable conditions
for successful shock revival than their FMD counterparts,
but the differences between successful and unsuccessful RbR+ simulations
(corresponding to high and low polar resolutions, respectively)
are caused by pole effects.

\begin{figure}
    \includegraphics[width=0.48\textwidth]{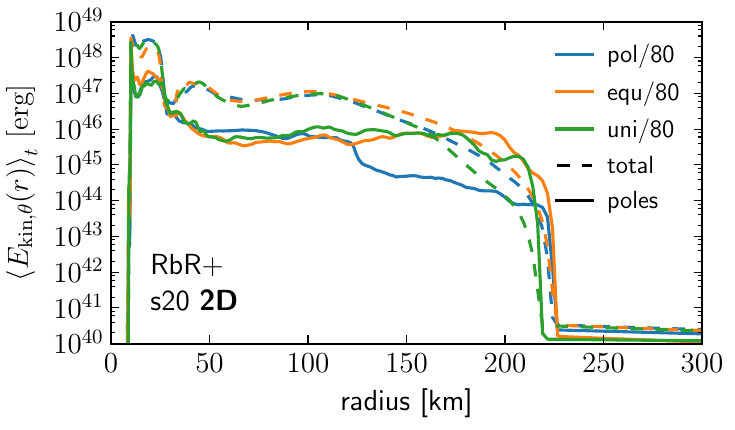}
    \caption{Lateral kinetic energies ${E_{\mathrm{kin}, \theta} (r)}$ for three different 2D RbR+ simulations of the s20 progenitor model.
            We show ${E_{\mathrm{kin}, \theta} (r)}$,
            averaged over a time interval ${t \in [300\unitspace\mathrm{ms}, 350\unitspace\mathrm{ms}]}$,
            as functions of radius $r$.
            The total lateral kinetic energies (dashed lines)
            are obtained by integrating the kinetic energy over spherical shells at each radius $r$
            (see Equation~\eqref{eqn:turbulence_energy_poles}).
            For polar kinetic energies (solid lines), the shell integral includes only regions close to the poles with
            polar angles ${\theta \in [0,10^{\circ}]}$ and ${\theta \in [170^{\circ}, 180^{\circ}]}$.
            To facilitate comparison, the polar energies are rescaled by the ratio of the total shell volume
            to the polar volume at each radius.
            Notice that the radial grids of all three simulations are identical,
            so we can compare integrals over radial shells in a straightforward manner.
            \label{fig:s20_turbulence_2D}}
\end{figure}

In order to understand the subtle differences that determine the explosion behavior of some 2D RbR+ models
while other RbR+ models fail to explode,
we consider the evolution of relevant quantities in the immediate vicinity of the poles.
We do not find any discriminative differences until ${350\unitspace\mathrm{ms}}$ p.b.\ in the polar neutrino luminosities and energies,
nor in the polar neutrino heating and cooling rates.
However, we find turbulent motions at the poles to be suppressed in models with coarse polar resolution.
This can be seen in Figure~\ref{fig:s20_turbulence_2D},
where we plot the lateral kinetic energies near the poles for three different 2D high-resolution simulations with the RbR+ approximation.
The lateral kinetic energy at a radius $r$ is obtained by integrating over the radial shell volume ${V_{\mathrm{shell}}(r)}$,
\begin{equation}\label{eqn:turbulence_energy_poles}
E_{\mathrm{kin}, \theta} (r) =
\int_{V_{\mathrm{shell}}(r)} \rho \; \frac{v_{\theta}^2}{2} \dint{V} ,
\end{equation}
with $v_{\theta}$ being the polar velocity component.
We differentiate between total lateral kinetic energies (dashed lines),
for which we integrate over the entire radial shell with polar angles ${\theta \in [0^{\circ}, 180^{\circ}]}$,
and lateral kinetic energies near the poles (solid lines),
for which the volume ${V_{\mathrm{shell}}(r)}$ includes only the regions of the radial shell
in the vicinity of the poles for polar angles
${\theta \in [0,10^{\circ}]}$ and ${\theta \in [170^{\circ}, 180^{\circ}]}$.
To facilitate comparison, we rescale the polar energies by the ratio of total shell volume to polar volume at each radius.
Figure~\ref{fig:s20_turbulence_2D} shows the lateral kinetic energies
averaged over the time interval ${t \in [300\unitspace\mathrm{ms}, 350\unitspace\mathrm{ms}]}$,
which corresponds to the phase of violent turbulent motions in the gain layer shortly before exploding models exhibit shock runaway.
In the region behind the shock, i.e.\ between ${120\unitspace\mathrm{km}}$ and ${220\unitspace\mathrm{km}}$,
the polar kinetic energies are significantly suppressed in the simulation with coarser polar grid resolution (`pol/80', solid blue line),
when compared to the other two simulations (solid orange and green lines),
but also when compared to the total kinetic energy (dashed blue line).
For simulations with high polar resolution (`equ/80' and `uni/80'),
the polar energies exceed the total energies (dashed orange and green lines) for radii between ${170\unitspace\mathrm{km}}$
and ${220\unitspace\mathrm{km}}$.

Since all of our successful explosions of the s20 progenitor model in 2D are driven by strong SASI sloshing motions in polar directions,
we suspect that in simulations with coarser angular resolution at the poles the suppression of turbulent motions
leads to conditions that are less favorable for a successful explosion because of weaker turbulent effects
(see, e.g., \citealt{2013ApJ...771...52M,2015MNRAS.448.2141M,2016ApJ...820...76R,2018ApJ...856...22M})
in particular around the grid axis.
For 2D simulations of the s20 model, this suffices to turn a successful explosion into a failed one.

\begin{figure}
    \includegraphics[width=0.48\textwidth]{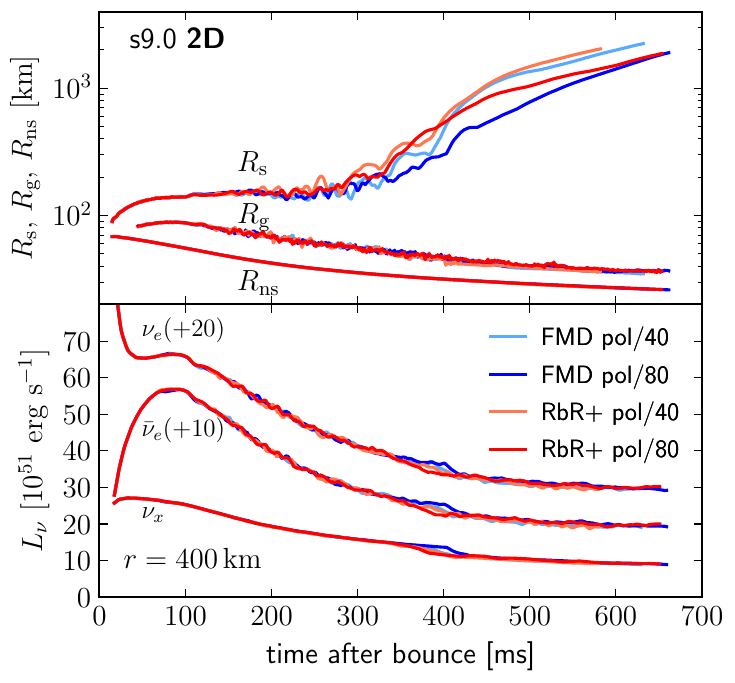}
    \caption{Comparison of 2D simulations of the s9.0 progenitor model.
            Similarly to Figure~\ref{fig:s20_comparison_2D}, we show, as functions of the time after bounce,
            the angle-averaged shock radii $R_{\mathrm{s}}$,
            gain radii $R_{\mathrm{g}}$, and neutron-star radii $R_{\mathrm{ns}}$ (all in the top panel; see labels),
            and the neutrino luminosities $L_{\nu}$ (bottom panel) for \nue, \nubar, and \nux (see labels).
            The neutrino luminosities are evaluated at a radius of ${r = 400\unitspace\mathrm{km}}$ in the co-moving frame of the stellar fluid.
            We differentiate the 2D simulations of the s9.0 model by color (see legend).
            Note that the luminosities were smoothed by running averages of ${5\unitspace\mathrm{ms}}$,
            and some lines were shifted vertically to facilitate readability of the plot.
            These lines are marked with labels indicating the number by which they are shifted in units of the ordinate.
            \label{fig:S90_comparison_2D}}
\end{figure}

In order to compare simulations with the RbR+ approximation and FMD neutrino transport in the case of successful CCSN explosions,
we chose the low-mass iron-core progenitor star s9.0 for our second set of 2D and 3D simulations
(compare \citealt{2017ApJ...850...43R} and \citealt{2018MNRAS.481.4786J},
who found explosions for the s9.0 model in 2D simulations with FMD neutrino transport).
For the 2D simulations of the s9.0 model, we restricted ourselves to the polar grids that we use in our 3D simulations
(`pol/40' and `pol/80').

In Figure~\ref{fig:S90_comparison_2D} we present an overview of all 2D simulations of the s9.0 progenitor model.
As the temporal evolution of the angle-averaged shock radii (top panel) shows,
all simulations exhibit shock revival within ${300\unitspace\mathrm{ms}}$ to ${400\unitspace\mathrm{ms}}$ after core bounce.
Apart from stochastic fluctuations, we find only minor differences in the evolution of the shock radii
and perfect agreement for gain radii and neutron-star radii.
Simulations with low grid resolution and with the RbR+ approximation tend to exhibit slightly larger shock radii,
when compared to simulations with high resolution and with FMD transport, respectively.
The time-dependent evolution of the neutrino properties agrees very well between all 2D simulations.
The bottom panel of Figure~\ref{fig:S90_comparison_2D} displays, as an example, the neutrino luminosities
(evaluated at a radius of ${r = 400\unitspace\mathrm{km}}$ in the co-moving frame),
which steadily fall due to the decline of mass accretion onto the neutron star.
The slightly more pronounced drop in the luminosities between ${350\unitspace\mathrm{ms}}$ and ${420\unitspace\mathrm{ms}}$
corresponds to Doppler-shift effects in the co-moving frame quantities
associated with the absolute increase of the radial velocity at $400\unitspace\mathrm{km}$ when the shock passes this radius.

In summary, we find our 2D simulations for both progenitor models, s20 and s9.0, to be consistent with results by other groups.
For a detailed assessment of 2D simulations of the s20 and s9.0 progenitor models with the \textsc{Alcar} code,
including comparisons to results from other codes, we refer the reader to \citet{2018MNRAS.481.4786J}.
We will come back to our 2D simulations in some parts of the next section when comparing results between 2D and 3D simulations.

\section{Three-Dimensional Simulations}\label{sec:three-dimensional}

In this section we present the results of our self-consistent, time-dependent CCSN simulations in three spatial dimensions.
We conducted four simulations for each progenitor model (s20 and s9.0),
and used either the RbR+ approximation or the FMD scheme for neutrino transport, respectively.
To assess the influence of the grid resolution, we performed all simulations with low (labeled with suffix ``L''; see Table~\ref{tab:grids})
as well as with high (i.e., twice the number of grid zones in every spatial direction; labeled with suffix ``H'') resolution.
Except for the neutrino-transport method (RbR+ or FMD) and the grid resolution,
all of our simulations are based on the same physical and numerical inputs, including neutrino opacities, the equation of state,
the treatment of gravity, and all other numerical details (see Section~\ref{sec:numerics} for the simulation setup).
For this reason, we can directly assess the influence of the neutrino-transport method on the delayed heating mechanism in 3D CCSN simulations.
To facilitate the presentation of our results, we divide this section into two subsections, each dealing with one of the progenitor models.

\subsection{Results for the s20 Progenitor Model}\label{sec:s20_3D}

\begin{figure}
    \includegraphics[width=0.48\textwidth]{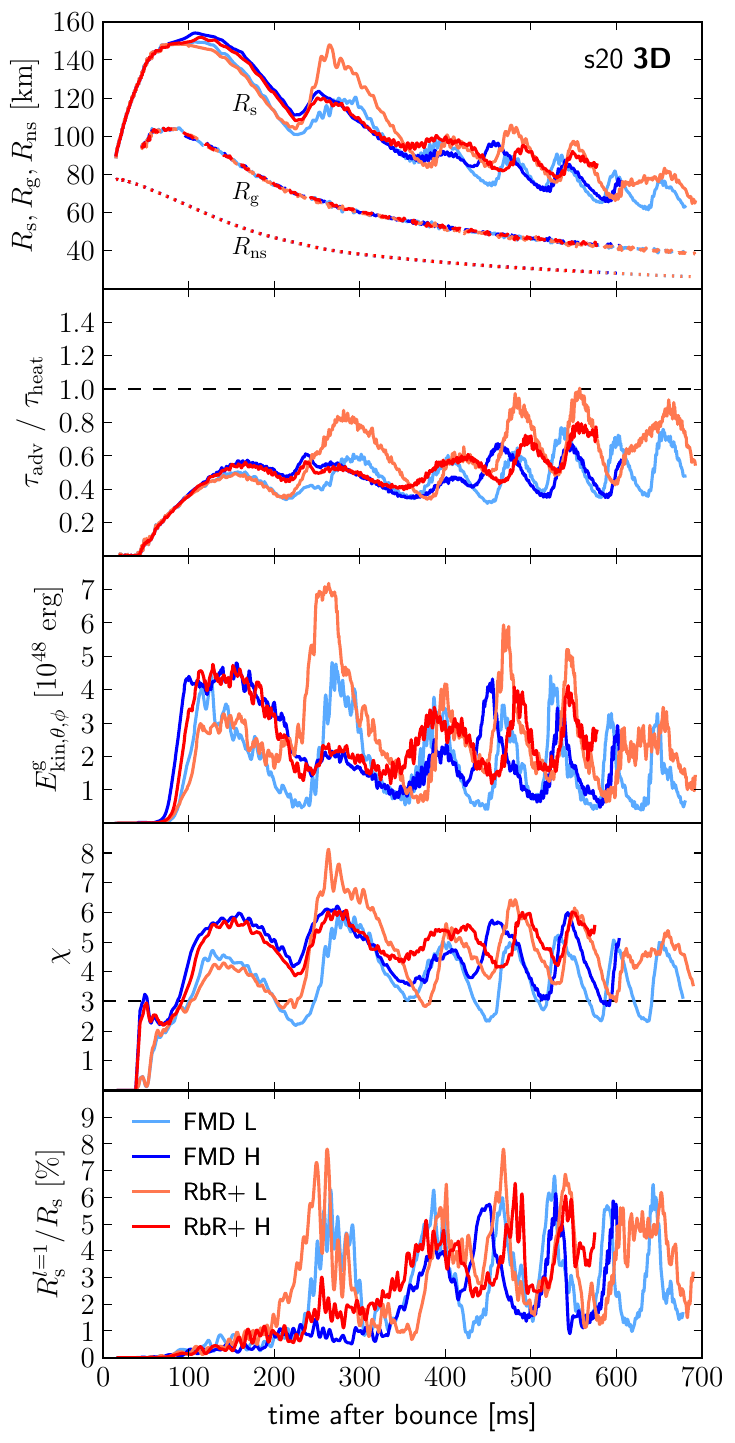}
    \caption{Overview of all 3D simulations of the s20 progenitor model.
            Shown are, as functions of the time after bounce, the angle-averaged shock radii $R_{\mathrm{s}}$ (solid lines),
            average gain radii $R_{\mathrm{g}}$ (dashed lines),
            and neutron-star radii $R_{\mathrm{ns}}$ (dotted lines; all in the first panel),
            the ratios of advection to heating timescales $\tau_{\mathrm{adv}} / \tau_{\mathrm{heat}}$ (second panel),
            the non-radial kinetic energies in the gain layer $E^{\mathrm{g}}_{\mathrm{kin},\theta,\phi}$ (third panel),
            the $\chi$ parameters (fourth panel), and the dipole moments of the angle-dependent shock radii $R_{\mathrm{s}}^{l=1}$,
            normalized to corresponding values of the monopole moments $R_{\mathrm{s}}$ (fifth panel).
            The different line colors correspond to low- (suffixes ``L'') and high-resolution (suffixes ``H'') simulations
            with either the FMD neutrino-transport scheme or the RbR+ approximation (see legend in the bottom panel).
            Note that $\chi$ and $R_{\mathrm{s}}^{l=1}$ were smoothed by running averages of ${5\unitspace\mathrm{ms}}$.
            \label{fig:radius_s20_3D}}
\end{figure}

We carried out four 3D simulations of the s20 progenitor model and followed the post-bounce phase for at least ${500\unitspace\mathrm{ms}}$.
For a general overview of our simulations, we present the temporal evolution of several diagnostic quantities in Figure~\ref{fig:radius_s20_3D}.
The first panel shows the angle-averaged shock radii $R_{\mathrm{s}}$ (solid lines),
the average gain radii $R_{\mathrm{g}}$ (dashed lines),
and the neutron-star radii $R_{\mathrm{ns}}$ (dotted lines),
which are defined as the radii at which the angle-averaged density drops below ${10^{11}\unitspace\gpercc}$.
Both neutron-star radii and gain radii show an essentially identical behavior in all four simulations (see legend in the last panel),
indicating that the influence of the neutrino-transport method and the grid resolution on the neutron-star contraction are negligible.

The evolution of the shock radii reveals various features that are present in all four simulations:
In the first ${100\unitspace\mathrm{ms}}$ after bounce, the stalled shock
(having experienced energy drain by nuclear photodisintegration and neutrino losses)
expands to roughly ${150\unitspace\mathrm{km}}$,
pushed outward by the accreted mass that assembles itself around the nascent neutron star.
Subsequently, the shock begins to retreat as the neutron star contracts in response to neutrino losses.
Shortly after ${200\unitspace\mathrm{ms}}$, the Si/Si-O interface falls through the shock,
accompanied by a steep decrease in the mass accretion rate (compare Figure~\ref{fig:comparison_1D_progenitors_mdot}),
which causes an almost immediate increase of the shock radius.
The subsequent evolution depends strongly on the influence of hydrodynamic instabilities,
in particular SASI mass motions,
which lead to alternating phases of shock expansion and shock retraction until the end of our simulations.
For the low-resolution simulations we find slightly stronger oscillations
and larger values of the shock radius in the case of the RbR+ approximation.
For the high-resolution simulations, however, the differences in the evolution of the shock radii between RbR+ and FMD are only minor and,
when disregarding fluctuations that arise due to the stochastic nature of hydrodynamic instabilities,
the RbR+ and FMD simulations are in very good agreement with each other.
Without any indications of shock runaway,
the average shock radii decrease to ${80\unitspace\mathrm{km}}$ or less at the end of all 3D simulations.
In contrast to some of the 2D cases (compare Figure~\ref{fig:s20_comparison_2D}),
we do not find any successful explosion in 3D simulations of the s20 model.

As already discussed in Section~\ref{sec:two-dimensional}, successful shock revival depends on the conditions in the gain layer.
Favorable conditions for an explosion arise when the advection timescale ${\tau_{\mathrm{adv}}}$ exceeds the heating timescale ${\tau_{\mathrm{heat}}}$,
resulting in a ratio ${\tau_{\mathrm{adv}} / \tau_{\mathrm{heat}} \gtrsim 1}$.
The second panel of Figure~\ref{fig:radius_s20_3D} shows the ratio of advection to heating timescale
for all 3D simulations of the s20 progenitor model.
Corresponding to phases of shock expansion and shock retraction,
all simulations exhibit phases with conditions that are more and less favorable for shock revival,
i.e.\ where  ${\tau_{\mathrm{adv}} / \tau_{\mathrm{heat}}}$ is close to unity, or much smaller than unity, respectively.
In low-resolution simulations, the RbR+ approximation leads to conditions that are more beneficial for shock revival
than results with FMD transport at most times.
This is consistent with the larger shock radii that we observe in the RbR+ simulation.
In contrast, the high-resolution simulations exhibit quite similar conditions for shock revival throughout the whole simulated evolution
with considerably smaller differences between RbR+ and FMD than in the low-resolution cases.

In addition to the aforementioned differences between simulations with RbR+ and FMD,
we find the high-resolution simulations to exhibit larger shock radii (ca.\ ${10\unitspace\mathrm{km}}$ difference)
between ${100\unitspace\mathrm{ms}}$ and ${220\unitspace\mathrm{ms}}$ after bounce, when compared to the low-resolution cases.
Consequently, the masses in the gain layer and, therefore, the advection timescales are larger in high-resolution simulations,
leading to higher timescale ratios in these cases.
These differences between low- and high-resolution simulations are caused by hydrodynamic instabilities in the post-shock region,
which are slightly stronger and, thus, push the shock further outward in the high-resolution simulations.
This can be seen from the third panel of Figure~\ref{fig:radius_s20_3D},
which shows the non-radial kinetic energies in the gain layer, ${E^{\mathrm{g}}_{\mathrm{kin},\theta,\phi}}$,
as functions of the time after bounce.
We calculate ${E^{\mathrm{g}}_{\mathrm{kin},\theta,\phi}}$ by the volume integral\footnote{Note
that Equations~\eqref{eqn:turbulence_energy_poles} and \eqref{eqn:ekin_lat_gain} define two different types of lateral kinetic energies.
Equation~\eqref{eqn:turbulence_energy_poles} (see Section~\ref{sec:two-dimensional}) defines ${E_{\mathrm{kin}, \theta} (r)}$ for radial shells,
whereas Equation \eqref{eqn:ekin_lat_gain} defines ${E^{\mathrm{g}}_{\mathrm{kin},\theta,\phi}}$ for the whole gain layer.}
\begin{equation}\label{eqn:ekin_lat_gain}
E^{\mathrm{g}}_{\mathrm{kin},\theta,\phi} =
\int_{V_{\mathrm{gain}}} \rho
\left( \frac{ v^{2}_{\theta} + v^{2}_{\phi} }{2} \right) \dint V ,
\end{equation}
with polar and azimuthal velocity components $v_{\theta}$ and $v_{\phi}$, respectively.
The high-resolution simulations exhibit a slightly earlier rise and larger values of the lateral kinetic energies until ${200\unitspace\mathrm{ms}}$ p.b.,
after which they start oscillating in all simulations in a similar manner as the angle-averaged shock radii.

A possible explanation for the different behavior between low- and high-resolution simulations
in these first ${200\unitspace\mathrm{ms}}$ is provided by  \citet{2009ApJ...697.1827F},
who show that higher (radial) grid resolution is beneficial for the growth of SASI,
and results from 3D CCSN simulations by \citet{2015MNRAS.452.2071F} reveal higher kinetic energies in the gain layer
for simulations with higher angular resolution (see Figure 7d there).
In order to assess the evolution of hydrodynamic instabilities in the post-shock layer in more detail and, in particular,
to distinguish between SASI-dominated and convection-dominated phases in our simulations,
we analyze two additional diagnostic quantities.

First, we consider the $\chi$ parameter \citep{2006ApJ...652.1436F},
which provides information about the conditions for the growth of convection in the post-shock layer.
Favorable conditions arise when buoyant mass motions can set in faster than seed perturbation get advected out of the gain layer,
i.e., when the timescale for the growth of convective activity is short compared to the advection timescale.
The $\chi$ parameter essentially measures the ratio of advection timescale to convective growth timescale in the gain layer,
\begin{equation}\label{eqn:chi_parameter}
\chi =
\int \frac{ \left< \omega_{\mathrm{BV}} \right> }
{\left| \left< v_r \right> \right|} \dint{r},
\end{equation}
with the Brunt-V\"ais\"al\"a frequency ${\omega_{\mathrm{BV}}}$ and the angle-averaged radial velocity ${\left< v_r \right>}$.
We apply linear averaging of $\omega_{\mathrm{BV}}$ as described in Appendix~\ref{sec:appendix_chi_parameter}\footnote{For
a detailed discussion of the various methods to calculate the $\chi$ parameter
and the definition of the Brunt-V\"ais\"al\"a frequency, also see Appendix~\ref{sec:appendix_chi_parameter}.},
and we only take into account regions in the gain layer with $R_\mathrm{g} < r < R_\mathrm{s} (\theta, \phi)$
that are locally unstable for convection, i.e.\ where ${\omega_{\mathrm{BV}} > 0}$.
According to \citet[][assuming the linear regime for the growth of perturbations]{2006ApJ...652.1436F},
a value of ${\chi \gtrsim 3}$ is necessary for the development of convective activity in the gain layer.

Short advection timescales (corresponding to small values of $\chi$) disfavor the growth of convection in the post-shock layer,
but they have been shown to amplify the linear growth rates of SASI
\citep[see, e.g.,][]{2006ApJ...652.1436F,2008A&A...477..931S,2012ApJ...761...72M}.
Thus, we can at least very roughly discriminate between conditions that are beneficial for convective growth (${\chi \gtrsim 3}$),
and conditions that favor the development of SASI activity ($\chi \lesssim 4$)\footnote{Note
that in addition to analyzing the $\chi$ parameter,
we inspected the time-dependent evolution of the entropy distribution to discriminate between phases of convection and SASI.
The sloshing and spiral modes of SASI can be identified by large-scale motions of the entropy-jump at the shock,
whereas the appearance of low- and high-entropy plumes indicates convective overturn.
We found that the $\chi$ parameter matches the phases of SASI growth ($\chi \lesssim 4$) and
of growth of convection (${\chi \gtrsim 3}$) quite well for all of our 3D simulations.}.

Second, we perform a multipole analysis and decompose the angle-dependent shock radii ${R_{\mathrm{s}} (\theta, \phi)}$
into real spherical harmonics ${Y^{m}_{l} (\theta, \phi)}$ of degree $l$ and order $m$ \citep[see, e.g.,][]{2012ApJ...759....5B,2013ApJ...768..115O}.
The multipole coefficients read \citep{2013ApJ...770...66H}
\begin{equation}\label{eqn:shock_radius_multipole_coefficients}
a^{m}_{l} =
\frac{(-1)^{|m|}}
{\sqrt{4 \pi (2 l + 1)}}
\int
R_{\mathrm{s}} (\theta, \phi)
Y^{m}_{l} (\theta, \phi)
\dint{\Omega} .
\end{equation}
With this normalization, the monopole moment (${l = 0}$) is trivially given by the angle-averaged shock radius,
${R^{l=0}_{\mathrm{s}} = a^{0}_{0}} = R_{\mathrm{s}}$,
and the dipole moment (${l = 1}$) is obtained by
\begin{equation}\label{eqn:shock_radius_dipole}
R^{l=1}_{\mathrm{s}} =
\sqrt{\sum_{m=-1}^{1} \left[ a^{m}_{1} (r) \right] ^{2}} .
\end{equation}
For later reference, we identify the components of the dipole vector in Cartesian coordinates to coincide with
${a_{x} = a^{1}_{1}}$, ${a_{y} = a^{-1}_{1}}$, and ${a_{z} = a^{0}_{1}}$.
The large-scale motions of SASI sloshing and spiral modes are reflected in large amplitudes of dipolar or quadrupolar character,
whereas convective overturn can both result in smaller-scale deformation of the shock
(i.e.\ excitation of higher $l$-modes, $l > 2$, of shock asymmetry)
and in large-scale deformation when large high-entropy plumes ascend toward the shock surface.

Figure~\ref{fig:radius_s20_3D} shows the temporal evolution of the $\chi$ parameters in panel four
and the dipole moments of the shock deformation, normalized by the angle-averaged shock radii, in panel five.
Starting at ${100\unitspace\mathrm{ms}}$ after bounce,
we find high-resolution simulations because of larger shock radii to exhibit more favorable conditions for convective growth,
i.e.\ systematically larger values of $\chi$ than in the low-resolution cases until ${250\unitspace\mathrm{ms}}$ after bounce.
The increasing dipole moments of the shock deformation reveal
that convective overturn in high-resolution simulations becomes more violent with time,
with the dipole moments reaching up to 3\% of the monopoles' amplitudes at ${\sim 250\unitspace\mathrm{ms}}$.
In contrast, the low-resolution simulations exhibit only weak convective activity with ${\chi \lesssim 4}$ and
stay SASI-dominated until at least ${250\unitspace\mathrm{ms}}$ after bounce.
They develop their first violent SASI activity with a combination of sloshing and spiral modes
between ${200\unitspace\mathrm{ms}}$ and ${300\unitspace\mathrm{ms}}$,
which cause the angle-averaged shock radii to exhibit large oscillations
and the dipole amplitudes of the shock deformation to reach up to 8\% of the average shock radii.
Due to the infall of the Si/Si-O interface shortly after ${200\unitspace\mathrm{ms}}$ p.b.,
the shock rapidly expands in both low- and high-resolution simulations,
leading to an increase of the mass in the gain layer and, consequently, of the advection timescale.
As a result, all simulations experience a convection-dominated phase
at around ${300\unitspace\mathrm{ms}}$ after bounce.

The subsequent evolution after ${300\unitspace\mathrm{ms}}$ p.b.\ is similar in all simulations
with alternating phases of convection and SASI, which represent repetitions of the following cycle:
During a shock-retraction phase, small shock radii and short advection timescales amplify the growth rates of SASI.
Combinations of SASI sloshing and spiral modes evolve and expand the shock front, leading to larger advection timescales,
which favor the growth of convective motions and damp the further growth of SASI.
With the shock reaching a maximal radius, the cycle enters a convection-dominated phase,
before the shock recedes again and SASI growth sets in once again.

\begin{figure}
    \includegraphics[width=0.48\textwidth]{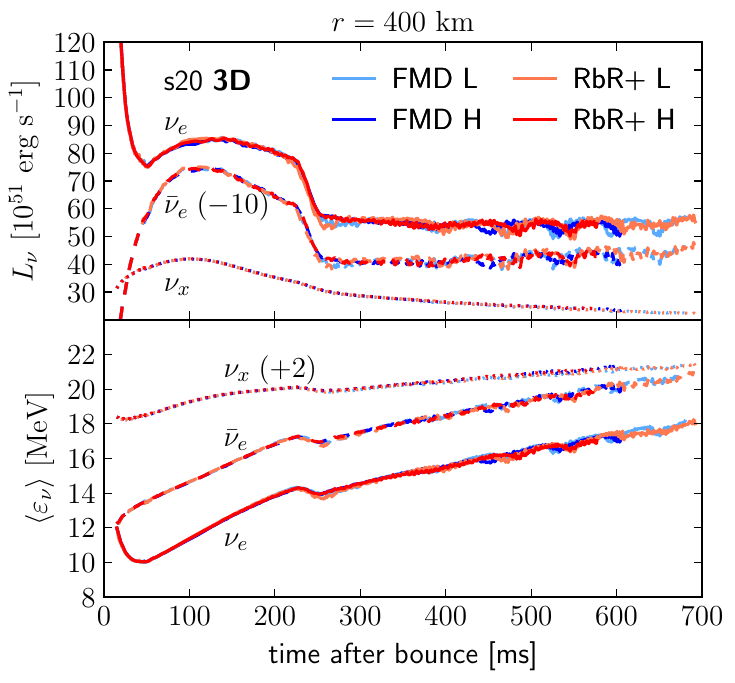}
    \caption{Neutrino luminosities $L_{\nu}$ (top panel)
            and neutrino mean energies $\left< \varepsilon_{\nu} \right>$ (bottom panel)
            as functions of time after bounce for all 3D simulations of the s20 progenitor model.
            Luminosities and mean energies are measured at a radius of ${r = 400\unitspace\mathrm{km}}$
            in the co-moving frame of the stellar fluid for electron neutrinos \nue (solid lines),
            electron antineutrinos \nubar (dashed lines), and heavy lepton neutrinos \nux (dotted lines).
            Note that some lines were shifted vertically to facilitate readability of the plot.
            These lines are labeled with the number by which they are shifted in units of the ordinate in parentheses.
            \label{fig:plot_s20_neutrinos_3D}}
\end{figure}

The evolution of hydrodynamic instabilities in the post-shock region with alternating phases of SASI and convection
and shock expansion and contraction influences the mass accretion rate onto the neutron star.
As a consequence, temporal fluctuations in the angle-averaged densities and temperatures occur at the surface of the neutron star.
Since electron-type neutrinos (\nue and \nubar) energetically decouple from the stellar matter in this region,
their luminosities and energies reflect the time-dependent fluctuations of the thermodynamic conditions at the neutron-star surface.
The top panel of Figure~\ref{fig:plot_s20_neutrinos_3D} shows corresponding fluctuations
of the order of 10\% to 20\% in both the \nue and \nubar luminosities
starting at ca.\ ${400\unitspace\mathrm{ms}}$ after bounce.
On a much smaller scale, these variations also exist in the mean energies of \nue and \nubar
(see bottom panel of Figure~\ref{fig:plot_s20_neutrinos_3D}).
The mean energies of the radial neutrino fluxes are defined for any species $\nu$ at a radius $r$ by
\begin{equation}\label{eqn:neutrino_mean_energy}
\left< \varepsilon_{\nu} (r) \right> =
\frac{\int F^{r}_{\nu} (r) \dint{\varepsilon} \dint{\Omega}}
{\int F^{r}_{\nu} (r) \; \varepsilon^{-1} \dint{\varepsilon} \dint{\Omega}} .
\end{equation}
The luminosities and mean energies of heavy-lepton neutrinos (\nux) are hardly affected by fluctuations in the density and temperature
at the neutron-star surface, because they are produced only much deeper inside the neutron star
by thermal pair creation and bremsstrahlung processes,
and scattering of \nux on electrons and nucleons does not suffice for \nux
to stay in thermodynamic equilibrium with the stellar fluid in the accretion layer.

\begin{figure*}
    \includegraphics[width=\textwidth]{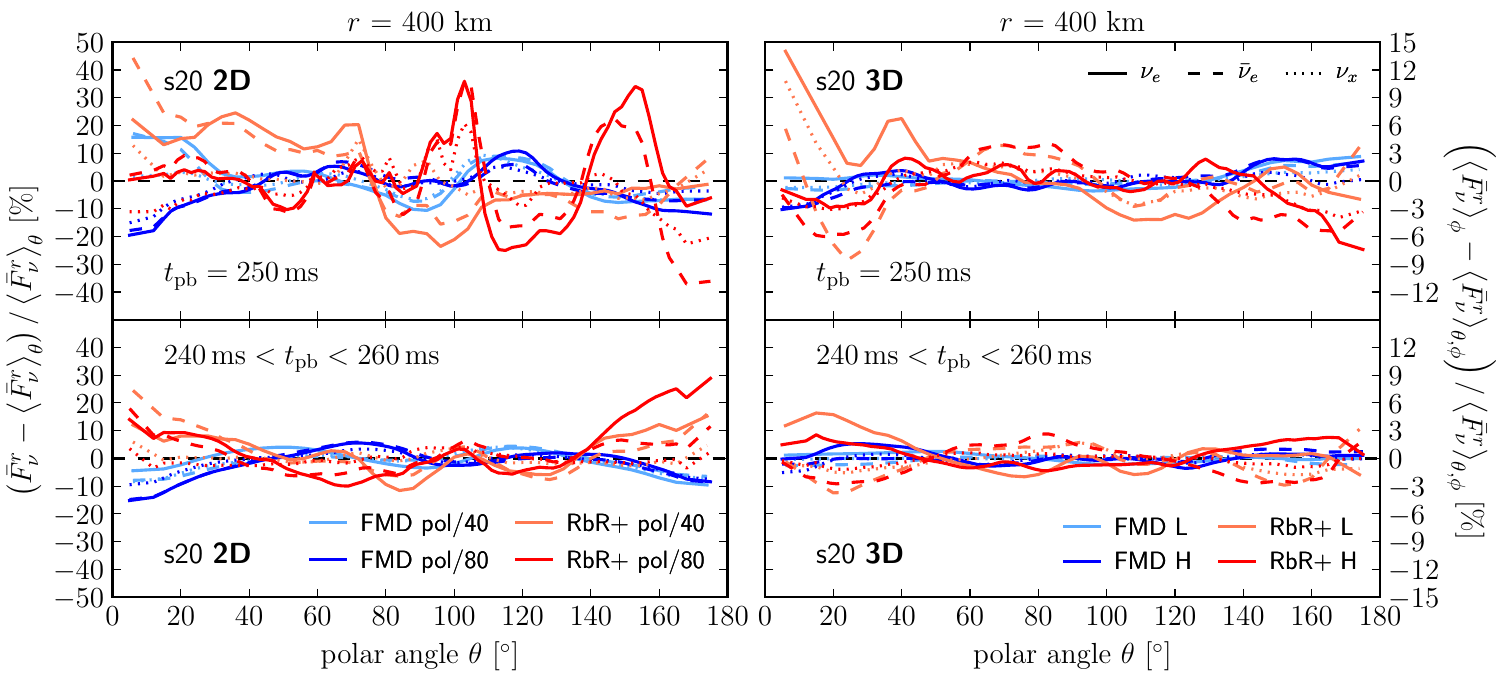}
    \caption{Lateral variations of energy-integrated radial neutrino fluxes ${\bar{F}^{r}_{\nu}}$,
            relative to their angular averages ${\left< \bar{F}^{r}_{\nu} \right>}$
            and normalized by the angle-averaged fluxes, as functions of the polar angle $\theta$
            for various simulations of the s20 progenitor model (see legends in the bottom panels, indicative for each column)
            for \nue (solid lines), \nubar (dashed lines), and \nux (dotted lines).
            We compare 2D simulations (left panels) and 3D simulations (right panels).
            For the latter the fluxes are averaged over the azimuthal angle $\phi$
            (i.e., ${\left< \bar{F}^{r}_{\nu} \right>_{\phi}}$ and ${\left< \bar{F}^{r}_{\nu} \right>_{\theta,\phi}}$, respectively).
            The top panels show flux deviations for temporal snapshots of each simulation
            at a post-bounce time of ${t_{\mathrm{pb}} = 250\unitspace\mathrm{ms}}$,
            whereas the bottom panels show time-averaged deviations in the time interval
            ${240\unitspace\mathrm{ms} < t_{\mathrm{pb}} < 260\unitspace\mathrm{ms}}$.
            All fluxes are measured at a radius of ${r = 400\unitspace\mathrm{km}}$ in the co-moving frame of the stellar fluid.
            Note that the range of the ordinate changes from left to right panels.
            \label{fig:fluxes_theta_s20_2D_3D}}
\end{figure*}

Because the evolution of hydrodynamic instabilities is very similar in all simulations
with only phase shifts in the cyclic shock expansion and contraction phases due to the stochastic nature of SASI and convection,
there are also no significant differences
between the neutrino luminosities and mean energies of all of our 3D simulations of the s20 progenitor.
In particular, although the dynamical evolution exhibits some differences
between low-resolution and high-resolution simulations in the first ${300\unitspace\mathrm{ms}}$ after bounce,
the neutrino properties still agree to an excellent degree.
We conclude that surface-integrated luminosities and mean energies of neutrinos are very robust
and do not sensitively depend on the neutrino-transport method or the numerical grid resolution.

\begin{figure*}
    \includegraphics[width=\textwidth]{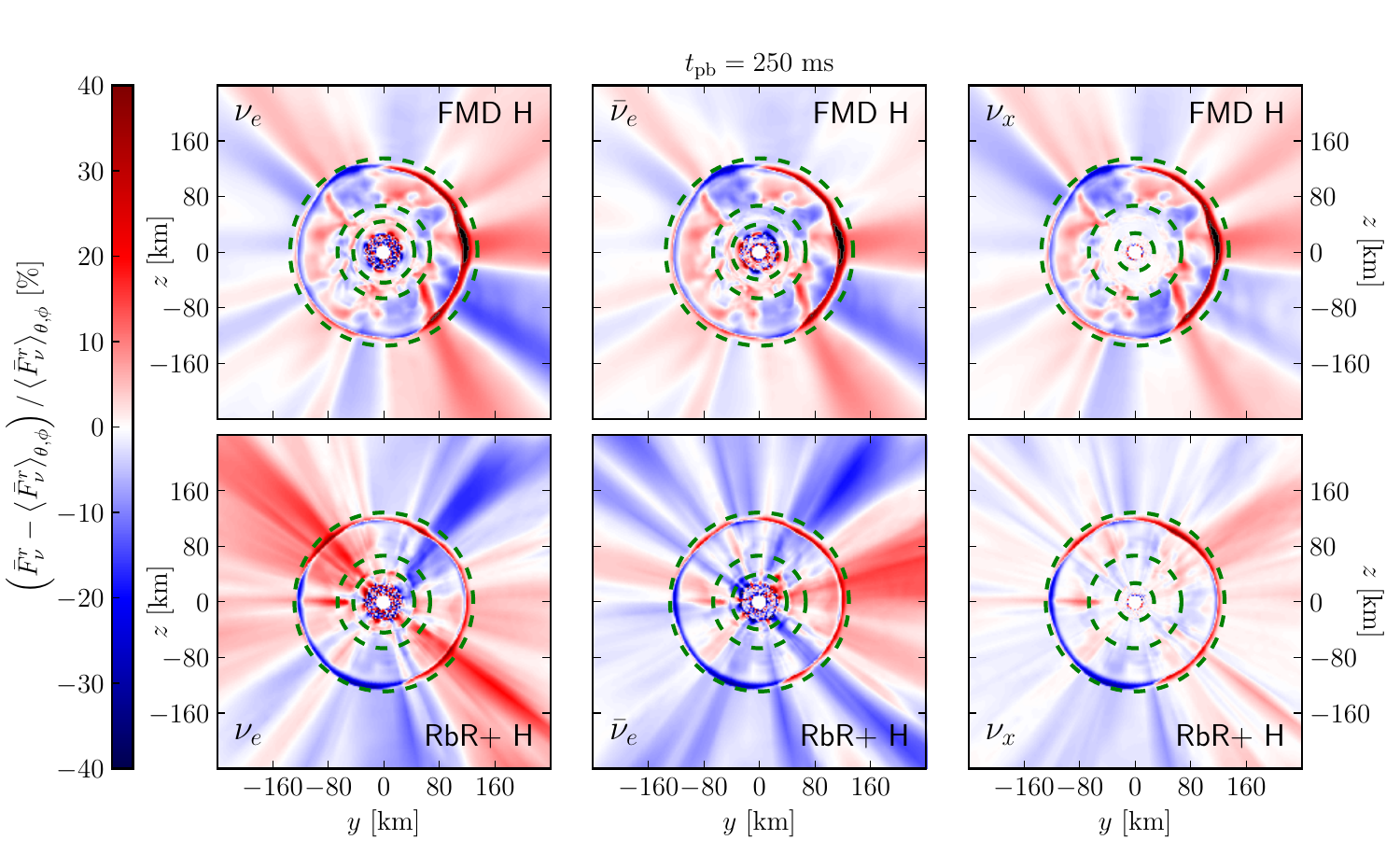}
    \caption{Relative variations between energy-integrated radial neutrino fluxes ${\bar{F}^{r}_{\nu}}$
        and their angular averages ${\left< \bar{F}^{r}_{\nu} \right>_{\theta,\phi}}$,
        normalized by the angle-averaged fluxes, for the 3D high-resolution simulations of the s20 progenitor model.
        We show the results in a plane ($y$-$z$) that contains the polar axis of our spherical polar grid.
        Results with FMD neutrino-transport scheme (top panels) are compared to the RbR+ approximation (bottom panels)
        for \nue (left panels), \nubar (central panels), and \nux (left panels).
        All fluxes are measured in the co-moving frame of the stellar fluid.
        Similar to Figure~\ref{fig:fluxes_theta_s20_2D_3D}, we show flux variations for temporal snapshots
        at a post-bounce time of ${t_{\mathrm{pb}} = 250\unitspace\mathrm{ms}}$.
        The inner (central, outer) green, dashed circles indicate the average neutrino spheres
        (average gain radii, maximum shock radii).
        Note that rings composed of dark red and dark blue parts just inside the maximum shock radii
        result from the discontinuous behavior of the co-moving radial fluxes at the angle-dependent shock radii.
        To facilitate comparisons, the color bar is limited to a range of $[-40\%, 40\%]$ and
        values outside of this interval are indicated by black color.
        \label{fig:slice_threedim_fluxes_s20_allnu}}
\end{figure*}

In addition to the analysis of surface-integrated luminosities, we also consider the angular variations of the neutrino fluxes.
Figure~\ref{fig:fluxes_theta_s20_2D_3D} shows the deviations of energy-integrated radial neutrino fluxes ${\bar{F}^{r}_{\nu}}$
to their angular averages ${\left< \bar{F}^{r}_{\nu} \right>}$ as functions of the polar angle $\theta$.
The energy-integrated radial fluxes for any neutrino species $\nu$ are obtained by ${\bar{F}_{\nu}^{r} = \int F^{r}_{\nu} \dint{\varepsilon}}$.
Since most variations of radial fluxes occur in phases of strong hydrodynamic instabilities,
we analyze the fluxes for a temporal snapshot at a post-bounce time of ${t_{\mathrm{pb}} = 250\unitspace\mathrm{ms}}$ (top panels),
when the low-resolution simulations experience strong shock expansion due to violent SASI activity,
and the high-resolution simulations exhibit strong convective overturn\footnote{Since we compare to 2D simulations,
we remark here that also the 2D simulations exhibit shock expansion caused by SASI sloshing motions in this phase
(see Section~\ref{sec:two-dimensional}).}.
The deviations of the radial fluxes from their angular averages are significantly larger in both 2D (left side)
and 3D (right side) simulations with the RbR+ approximation, when compared to the FMD neutrino-transport scheme.
This observation also holds for flux variations that have been averaged over a time interval of
${240\unitspace\mathrm{ms} < t_{\mathrm{pb}} < 260\unitspace\mathrm{ms}}$ (see bottom panels).
However, the amplitude of variations is significantly smaller in 3D simulations than in 2D models, e.g.,
it is less than 15\% for RbR+ and 3\% for FMD results for the temporal snapshot in 3D
compared to up to 45\% for RbR+ and 20\% for FMD for the 2D case
(notice that the range of the ordinate changes from left to right panels).
The time-averaged deviations decrease to 5\% or less in all 3D simulations,
although in particular the low-resolution RbR+ simulation exhibits strong SASI sloshing and spiral motions during this phase.
The time-averaged deviations reach up to 30\% in 2D simulations with the RbR+ approximation,
with the largest excursions for all neutrino species appearing at the poles
(where ${\theta \approx 0^{\circ}}$ and ${\theta \approx 180^{\circ}}$, respectively).
This observation strengthens our argument from Section~\ref{sec:two-dimensional},
where we suspected that in 2D simulations the RbR+ approximation
together with axial sloshing motions of SASI can amplify local variations at the poles
which are already visible in the simulations with FMD transport.
In contrast, we do not find this behavior in 3D simulations with RbR+
where time-averaged results of RbR+ and FMD models are much more similar
(see bottom right panel of Figure~\ref{fig:fluxes_theta_s20_2D_3D}).
We further remark that we do not observe any significant dependence of angular variations of radial fluxes on the grid resolution,
neither in 2D, nor in 3D simulations.

\begin{figure*}
    \includegraphics[width=\textwidth]{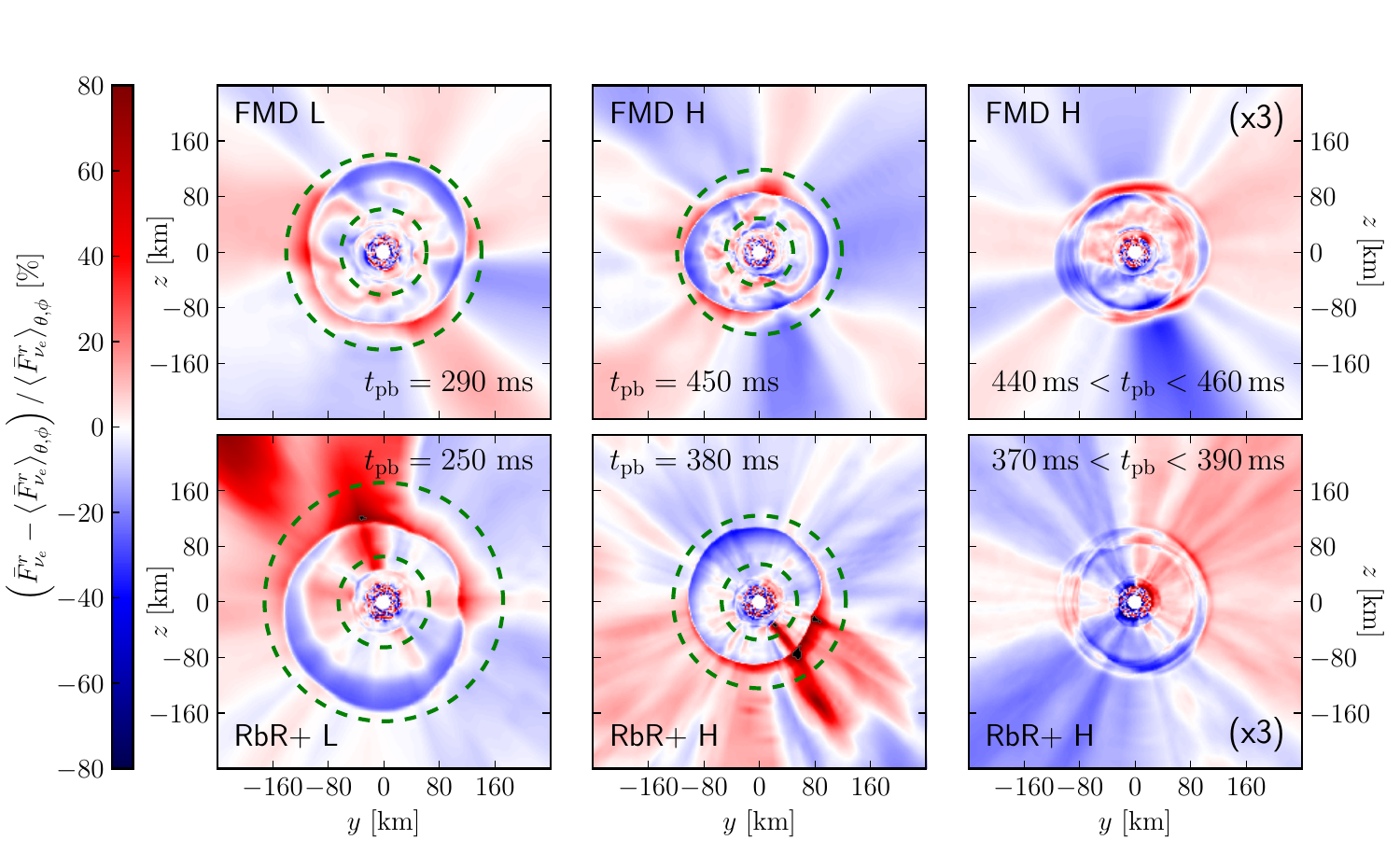}
    \caption{Relative neutrino-flux variations as in Figure~\ref{fig:slice_threedim_fluxes_s20_allnu},
            but only for electron neutrinos, and for both low-resolution (left panels) and high-resolution (center and right panels)
            simulations of the s20 progenitor model at temporal snapshots that differ from Figure~\ref{fig:slice_threedim_fluxes_s20_allnu}.
            The corresponding post-bounce times $t_{\mathrm{pb}}$ are noted in each panel.
            The flux variations in the right panel are averaged over time intervals of ${20\unitspace\mathrm{ms}}$.
            The inner (outer) green, dashed circles indicate the average gain (maximum shock) radii, respectively.
            To facilitate comparisons, the color bar is limited to a range of $[-80\%, 80\%]$ and
            values outside of this interval are indicated by black color.
            The time-averaged values in the right panels were scaled by a factor of 3.
            \label{fig:slice_threedim_fluxes_s20_times}}
\end{figure*}

In simulations with the RbR+ approximation, the non-radial flux components ${F^{\theta}_{\nu}}$ and ${F^{\phi}_{\nu}}$ are not evolved.
One consequence of this treatment can be seen in Figure~\ref{fig:slice_threedim_fluxes_s20_allnu},
which displays the deviations of ${\bar{F}^{r}_{\nu}}$ from the angular averages ${\left< \bar{F}^{r}_{\nu} \right>_{\theta,\phi}}$
for \nue (left), \nubar (center), and \nux (right)
with color coding in a plane ($y$-$z$) that contains the polar axis of the spherical polar grid.
The high-resolution RbR+ simulation (bottom panels) shows radial ``streaks'' for all neutrino species between the shock and
the neutrino energy sphere, at which neutrinos decouple energetically from the stellar matter,
i.e.\ where the energy-averaged effective optical depth fulfills ${\tau (r) = 2 / 3}$.
The optical depth is defined as
\begin{equation}\label{eqn:neutrino_sphere_effective}
\tau (r) = \int_{r}^{\infty} \kappa_{\mathrm{eff}} (r') \dint{r'},
\end{equation}
with $\kappa_{\mathrm{eff}} = \sqrt{\kappa_{\mathrm{abs}}\kappa_{\mathrm{tot}}}$ the effective opacity
calculated from the absorption and total opacities, $\kappa_{\mathrm{abs}}$ and $\kappa_{\mathrm{tot}}$,
and energetically averaged as in \citet{2006A&A...447.1049B}.
For \nue, e.g., the region between neutrino sphere (inner green, dashed circle)
and maximum shock radius (outer green, dashed circle)
approximately extends from ${44\unitspace\mathrm{km}}$ to ${129\unitspace\mathrm{km}}$.
In contrast, local variations in the radial neutrino fluxes are smoothed out by lateral flux components
in the simulation with FMD neutrino-transport scheme (top panels),
leading to a ``patch-like'' structure in the region between the neutrino sphere and the maximum shock radius
(e.g., between ${44\unitspace\mathrm{km}}$ and ${135\unitspace\mathrm{km}}$ for \nue).
The FMD simulation exhibits a ``streak-like'' pattern, as it is seen in the RbR+ case, only outside this region,
where non-radial neutrino flux components are approaching zero and, therefore, both neutrino-transport methods behave identically.
Apart from these observations, we find maximal flux deviations for \nue and \nubar
to be of the same order between RbR+ and FMD simulations,
and for \nux to be even smaller for the RbR+ simulation when compared to the FMD case.

\begin{figure*}
    \includegraphics[width=\textwidth]{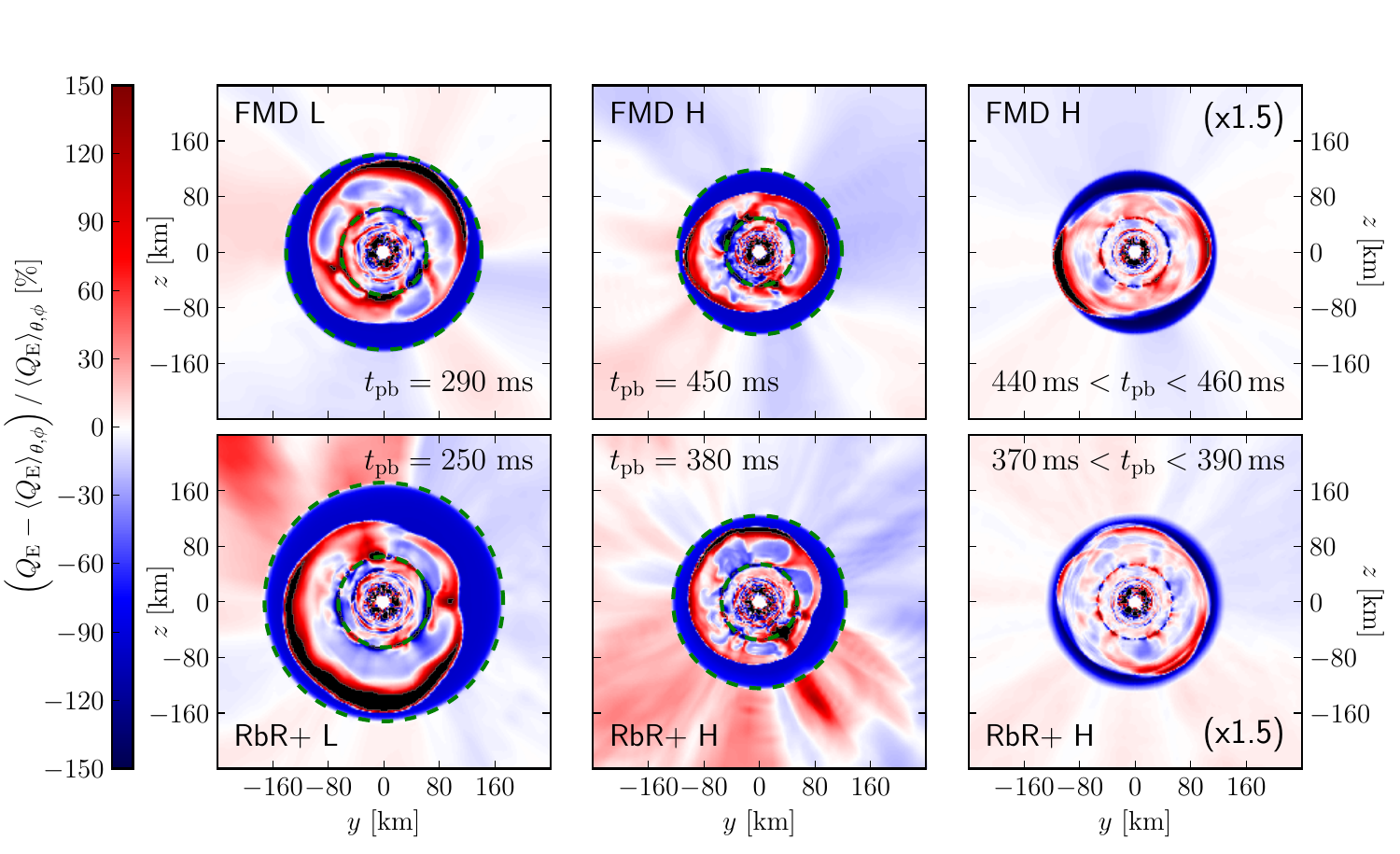}
    \caption{Relative variations between the local net neutrino heating rates $Q_{\mathrm{E}}$
            and their angular averages ${\left< Q_{\mathrm{E}} \right>_{\theta,\phi}}$,
            normalized by the angle-averaged values, for our 3D simulations of the s20 progenitor model.
            Correspondingly to Figure~\ref{fig:slice_threedim_fluxes_s20_times},
            the panels show data for temporal snapshots (left and center) and time-averaged values (right)
            in the same $y$-$z$-plane and for the same selection of simulations.
            The inner (outer) green, dashed circles indicate the average gain (maximum shock) radii, respectively.
            To facilitate comparisons, the color bar is limited to a range of $[-150\%, 150\%]$ and
            values outside of this interval are indicated by black color.
            The time-averaged values in the right panels were scaled by a factor of 1.5.
            Dark blue regions in the vicinity of the maximum shock radii are outside the gain layer and, thus,
            do not experience strong neutrino heating, whereas dark red and black regions are still inside the gain layer.
            Variations outside the maximum shock radii are irrelevant since the (absolute) heating rates are very small in these regions.
            \label{fig:slice_threedim_heating_s20}}
\end{figure*}

Previous findings from 2D simulations suggest that the RbR+ approximation can amplify local deviations in particular in combination
with strong axial sloshing activity of the SASI \cite[see, e.g.,][]{2016ApJ...831...81S,2018MNRAS.481.4786J}.
For this reason, we further investigate fluctuations of radial fluxes in our 3D simulations in phases of violent SASI mass motions.
Figure~\ref{fig:slice_threedim_fluxes_s20_times} again shows variations of ${\bar{F}^{r}_{\nue}}$ relative to the angular averages,
but for low- (left panels) and high-resolution (central panels) simulations in phases of strong SASI-activity
(the corresponding post-bounce times $t_{\mathrm{pb}}$ are noted in each panel).
We restrict ourselves to analyze only fluxes of \nue, since Figure~\ref{fig:slice_threedim_fluxes_s20_allnu}
revealed no fundamental differences in the behavior of \nue and \nubar
and relatively less prominent differences between RbR+ and FMD results for \nux.
In both simulations with the RbR+ approximation we find local ``hot spots'' at which the radial neutrino fluxes are high and,
thus, significantly deviate from the angular averages.
These deviations reach up to 80\% and more, e.g.\ in the top left sector for the low-resolution RbR+ simulation (bottom left panel),
and in the bottom right sector for the high-resolution RbR+ simulation (bottom center panel).
In contrast, the FMD simulations (upper left and  center panels) exhibit much lower variations.
The time-averaged deviations of ${\bar{F}^{r}_{\nue}}$ are much smaller
than corresponding instantaneous values for both high-resolution simulations,
e.g., when averaged over a ${20\unitspace\mathrm{ms}}$ time interval
(right panels; the values in the right panels are scaled up by a factor of 3),
and are more similar between high-resolution RbR+ and FMD simulations.

In order to estimate the influence of locally amplified neutrino fluxes as seen in the RbR+ simulations
on the neutrino heating in the gain layer, we consider deviations of the local net neutrino heating rates,
which are given by the energy source term $Q_{\mathrm{E}}$, from their angular averages ${\left< Q_{\mathrm{E}} \right>_{\theta,\phi}}$.
Figure~\ref{fig:slice_threedim_heating_s20} presents such deviations for low- and high-resolution 3D simulations
for the same temporal snapshots as in Figure~\ref{fig:slice_threedim_fluxes_s20_times},
i.e., during phases of violent SASI activity in all simulations.
The amplitudes of fluctuations of the local heating rates are of the same order in all 3D simulations.
Dark blue regions in the vicinity to the maximum shock radii are outside the gain layer and, therefore,
do not experience strong neutrino heating, whereas dark red and black regions are still inside the gain layer.
The spatial scales of regions with locally enhanced or reduced neutrino heating are large in low-resolution simulations,
whereas high-resolution simulations resolve finer structures in the gain layer.
However, the results do not show any significant differences in the spatial scales between simulations with RbR+ and FMD.
Furthermore, the time-averaged heating rates, which are scaled up by a factor of 1.5,
reveal very similar patterns in RbR+ and FMD simulations.
Regions with enhanced time-averaged heating rates (dark red and black regions) at a radius of ca.\ ${100\unitspace\mathrm{km}}$
to ${120\unitspace\mathrm{km}}$ indicate maximal excursions of the shock radii due to SASI motions.

\begin{figure*}
    \includegraphics[width=\textwidth]{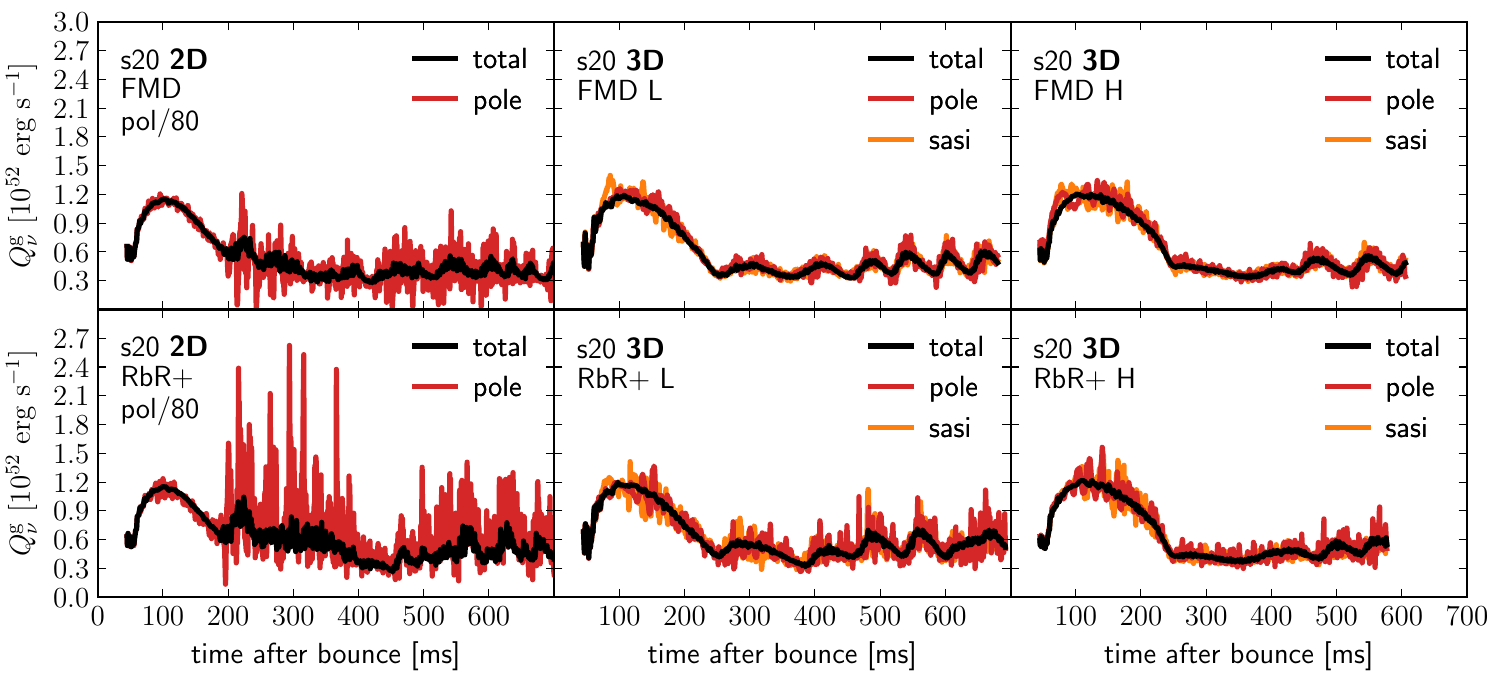}
    \caption{Net neutrino heating rates, integrated over the gain layer, ${Q^{\mathrm{g}}_{\nu}}$ (see Equation~\eqref{eqn:heating_gain_layer}),
            as functions of the time after bounce, for 2D and 3D simulations of the s20 progenitor model.
            2D simulations are shown in the left panels, 3D simulations with low-resolution in the central panels,
            and high-resolution results in the right panels.
            The top (bottom) panels display simulations with the FMD neutrino-transport scheme (with the RbR+ approximation), respectively.
            Total heating rates (labeled ``total''; black lines) are compared to
            the heating rates in a ${19^{\circ}}$ cone around the north pole (i.e., where ${\theta = 0^{\circ}}$; labeled ``pole''; red lines),
            and to the heating rates in a ${19^{\circ}}$ cone around the instantaneous SASI dipole direction (labeled ``sasi''; orange lines),
            respectively.
            To facilitate comparisons, the heating rates at the poles and in the SASI directions are rescaled
            by the ratios of the total volume of the gain layer to the volumes of the constrained cones of integration.
            \label{fig:heating_s20_2D_3D}}
\end{figure*}

In addition to instantaneous and time-averaged neutrino heating rates,
we further investigate the time-dependent evolution of the total (net) neutrino heating rates in the gain layer, ${Q^{\mathrm{g}}_{\nu}}$.
In Figure~\ref{fig:heating_s20_2D_3D} ${Q^{\mathrm{g}}_{\nu}}$ is compared for 2D simulations (left panels)
and 3D simulations with low (central panels) and high resolution (right panels), respectively.
The total heating rates (black lines),
which result from integrating the local energy source terms ${Q_{\mathrm{E}}}$ over the entire volume of the gain layer
(see Equation~\eqref{eqn:heating_gain_layer}), agree to high degree between all 2D and 3D simulations.
In contrast, a different behavior is observed for the polar heating rates (red lines),
for which the integral was carried out only over regions of the gain layer in the vicinity of the north pole,
i.e., where the polar angle ${\theta \in [0,19^{\circ}]}$.
To facilitate comparison, the heating rates at the pole are rescaled by the ratio of total volume to polar volume.
In the 2D simulation with the RbR+ approximation (lower left panel) the polar heating rates are significantly larger than in
the corresponding 2D simulation with FMD neutrino transport (upper left panel).
Furthermore, the polar heating rates are enhanced in comparison to the total rates in the 2D RbR+ simulation,
whereas they fluctuate around the total rates in the 2D FMD simulation with considerably lower amplitudes.
These findings support our argument, that in 2D simulations the RbR+ approximation overestimates polar variations in feedback with
strong axial SASI sloshing motions and, thus, it leads to conditions that are more favorable for successful shock revival
than those obtained in 2D simulations with an FMD neutrino-transport scheme.

However, in 3D simulations with FMD and RbR+ the polar heating rates show much less differences.
The low-resolution simulation with RbR+ exhibits slightly larger polar heating rates (central panels),
whereas the differences between FMD and RbR+ almost vanish for high-resolution simulations (right panels).
In general, we find the polar heating rates in 3D simulations to fluctuate much less around the total rates.
Since, in contrast to 2D simulations, SASI sloshing and spiral modes are not restricted to a prescribed direction in 3D simulations,
we also analyze the heating rates in the gain layer within a cone of half-opening angle of ${19^{\circ}}$
around the instantaneous direction of
the SASI mass motions, i.e., the dipole direction of the shock deformation, see Equations~\eqref{eqn:shock_radius_multipole_coefficients}
and \eqref{eqn:shock_radius_dipole}, and the associated explanations.
The neutrino heating rates near the SASI direction (orange lines; we again rescale the heating rates by the ratio of total volume to cone volume)
exhibit a very similar behavior as the polar heating rates,
with the RbR+ simulations showing only insignificantly larger heating rates in the SASI direction than the corresponding FMD simulations,
in particular in the high-resolution 3D simulations.

In summary, we find only minor differences between 3D simulations of the s20 progenitor model
with the RbR+ approximation and an FMD neutrino-transport scheme,
in particular when considering local deviations of neutrino fluxes and heating rates from their angular averages
in the case of time integration over periods of typically ${10\unitspace\mathrm{ms}}$ or longer
(see, e.g., Figures~\ref{fig:fluxes_theta_s20_2D_3D},
\ref{fig:slice_threedim_fluxes_s20_times}, and \ref{fig:slice_threedim_heating_s20}).
Moreover, surface-integrated fluxes and mean energies of neutrinos show excellent agreement (see Figure~\ref{fig:plot_s20_neutrinos_3D}),
and also the overall post-bounce evolution with alternating phases of dominant SASI or convective activity is similar
in all of our 3D simulations of the s20 progenitor model (see Figure~\ref{fig:radius_s20_3D}).
Considering gain-layer integrated neutrino-heating rates, we find only minor differences between FMD and RbR+
(see Figure~\ref{fig:heating_s20_2D_3D}),
which contrasts results from 2D simulations, in which polar heating rates are significantly amplified in the case of RbR+.

\subsection{Results for the s9.0 Progenitor Model}\label{sec:s90_3D}

\begin{figure}
    \includegraphics[width=0.48\textwidth]{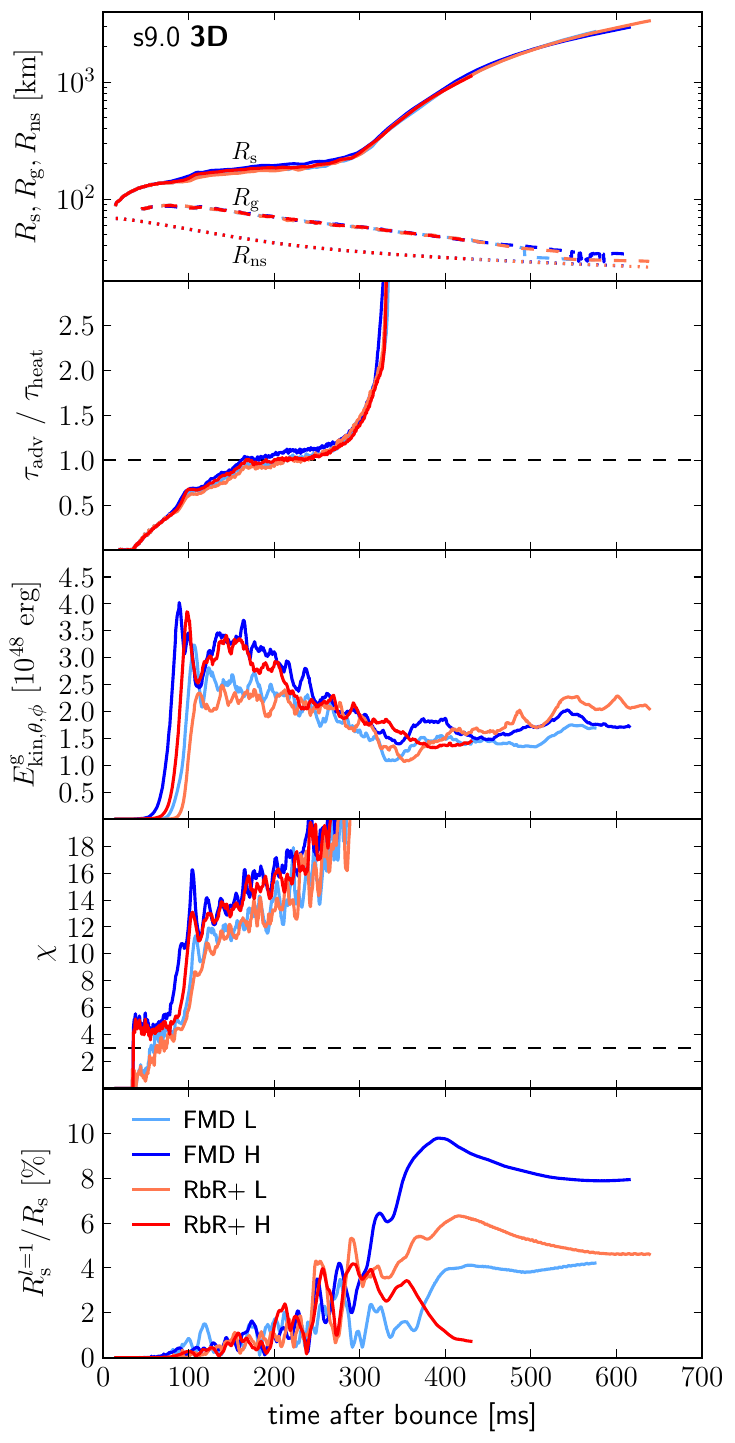}
    \caption{Overview of all 3D simulations of the s9.0 progenitor model.
        Shown are, as functions of the time after bounce, the angle-averaged shock radii $R_{\mathrm{s}}$ (solid lines),
        gain radii $R_{\mathrm{g}}$ (dashed lines), and neutron-star radii $R_{\mathrm{ns}}$ (dotted lines; all in the first panel),
        the ratios of advection to heating timescale $\tau_{\mathrm{adv}} / \tau_{\mathrm{heat}}$ (second panel),
        the non-radial kinetic energies in the gain layer $E^{\mathrm{g}}_{\mathrm{kin},\theta,\phi}$ (third panel),
        the $\chi$ parameters (fourth panel), and the dipole moments of the angle-dependent shock radii $R_{\mathrm{s}}^{l=1}$,
        normalized to corresponding values of the monopole moments $R_{\mathrm{s}}$ (fifth panel).
        The different line colors correspond to low- (suffix ``L'') and high-resolution (suffix ``H'') simulations
        with either the FMD neutrino-transport scheme or the RbR+ approximation (see legend in the bottom panel).
        \label{fig:radius_S90_3D}}
\end{figure}

In this section we discuss the results of our 3D simulations of the s9.0 progenitor model.
As in our 3D simulations of the s20 model, we used either the RbR+ approximation or the FMD scheme for neutrino transport,
and performed simulations with low as well as high grid resolution.
For a general overview, the temporal evolution of a selection of several diagnostic quantities is presented in Figure~\ref{fig:radius_S90_3D}.
We find successful shock revival at roughly ${300\unitspace\mathrm{ms}}$ after bounce for all four simulations,
which can be seen from the evolution of the angle-averaged shock radii (first panel)
and the ratios of advection timescale to heating timescale (second panel), which considerably exceed 1 at that time.
The agreement in the evolution of the average positions of shock, gain radius\footnote{At late times
(i.e., after about ${500\unitspace\mathrm{ms}}$),
the exact position of the boundary between heating and cooling layers sensitively depends on
highly time-variable local downflows onto the neutron star.
For this reason, the gain radii start to slightly differ between simulations in that phase.},
neutron-star radius, and of the timescale ratios
is excellent between all four 3D simulations of the s9.0 progenitor model (first and second panels).

Similar to our findings for 3D simulations of the s20 model (see Figure~\ref{fig:radius_s20_3D}),
we observe an earlier rise and larger values of the non-radial kinetic energies in the gain layer for the high-resolution simulations (third panel).
Because of the steep density gradient at the edge of the degenerate core in the s9.0 progenitor model
(see Figure~\ref{fig:comparison_1D_progenitors})
and the resulting low mass advection rate (see Figure~\ref{fig:comparison_1D_progenitors_mdot}),
we find relatively long advection timescales and, thus, favorable conditions for the growth of convection in all simulations of the s9.0 model.
Correspondingly, all simulations exhibit $\chi$ parameters considerably larger than the critical value of 3
starting at roughly ${100\unitspace\mathrm{ms}}$ after bounce (fourth panel).
For this reason, all simulations of the s9.0 progenitor model stay convection-dominated
and do not exhibit any obvious SASI activity until the end of the simulations.
Remarkably, we observe the dipole moments of the shock deformation to reach amplitudes up to 10\% of the monopole moments (fifth panel),
triggered mainly by large-scale convective plumes in the post-shock layer.
All in all, we do not find any major differences between all four 3D simulations of the s9.0 progenitor model,
neither between low- and high-resolution simulations, nor between RbR+ and FMD neutrino transport.

\begin{figure}
    \includegraphics[width=0.48\textwidth]{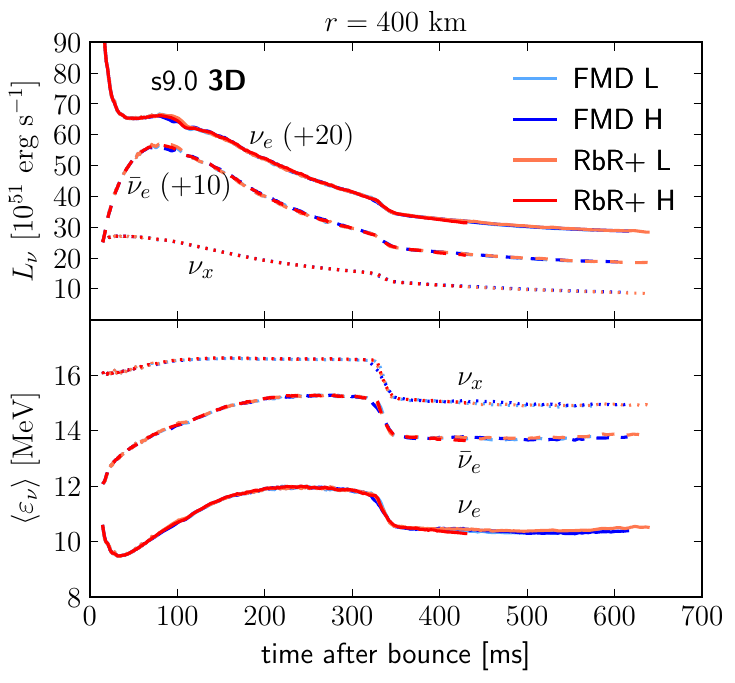}
    \caption{Neutrino luminosities $L_{\nu}$ (top panel)
            and neutrino mean energies $\left< \varepsilon_{\nu} \right>$ (bottom panel)
            as functions of the time after bounce for all 3D simulations of the s9.0 progenitor model.
            Luminosities and mean energies are measured at a radius of ${r = 400\unitspace\mathrm{km}}$
            in the co-moving frame of the stellar fluid.
            Different line styles show electron neutrinos \nue (solid lines), electron antineutrinos \nubar (dashed lines)
            and heavy lepton neutrinos \nux (dotted lines).
            To facilitate readability of the plot some lines were shifted vertically
            indicated by the numbers giving the shifts in units of the ordinate.
            \label{fig:plot_S90_neutrinos_3D}}
\end{figure}

Figure~\ref{fig:plot_S90_neutrinos_3D} shows the co-moving frame neutrino luminosities (top panel) and mean energies (bottom panel)
for all three neutrino species at a radius of ${r = 400\unitspace\mathrm{km}}$.
Again, the agreement between all four simulations is excellent.
Shortly after shock expansion at about ${300\unitspace\mathrm{ms}}$ p.b.,
both neutrino luminosities and mean energies significantly drop due to the decline of the mass accretion onto the neutron star.
As a result of continued cooling of the hot proto-neutron star,
both neutrino luminosities and mean energies exhibit almost constant values (decreasing only slowly with time) until the end of our simulations.

Compared to our 2D simulations of the s9.0 model (see Section~\ref{sec:two-dimensional} and Figure~\ref{fig:S90_comparison_2D}),
the 3D models exhibit a slightly earlier drop of the neutrino luminosities,
i.e., at ${\sim 320\unitspace\mathrm{ms}}$ p.b.\ instead of ${\sim 380}$ -- ${420\unitspace\mathrm{ms}}$ p.b.\ in 2D simulations,
resulting from slightly earlier shock expansion and a larger shock radius in 3D,
which correlate with initially higher turbulent kinetic energies in the gain layer pushing the shock farther out.
Furthermore, the shock radii show larger variations between simulations in 2D,
whereas 3D simulations display a very good agreement.

\begin{figure}
    \includegraphics[width=0.48\textwidth]{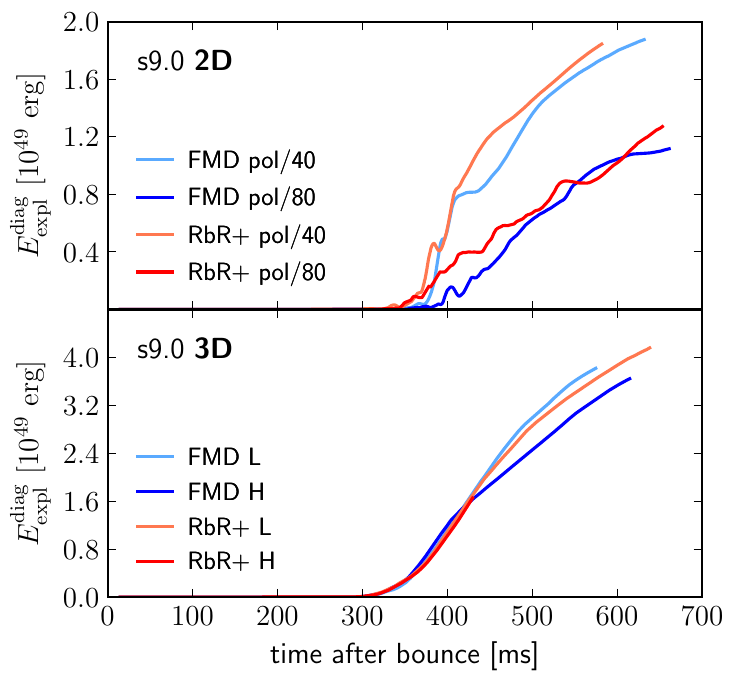}
    \caption{Diagnostic explosion energies $E^{\mathrm{diag}}_{\mathrm{expl}}$ in dependence on the time after bounce
            for all 2D (top panel) and 3D (bottom panel) simulations of the s9.0 progenitor model.
            The range of the ordinate differs between top and bottom panels.
            \label{fig:explosion_energies_s90_2D_3D}}
\end{figure}

Additional differences between 2D and 3D simulations of the s9.0 progenitor model appear in the diagnostic explosion energies.
These are defined as the total energy in the gain layer (see Equation~\eqref{eqn:total_energy_gain}),
\begin{equation}\label{eqn:diagnostic_explosion_energy}
E_{\mathrm{expl}}^{\mathrm{diag}} =
\int_{V_{\mathrm{ps}}}
\rho \left( e_{\mathrm{int}} + \frac{v^{2}}{2} + \phi_{\mathrm{grav}} \right) \dint V ,
\end{equation}
with the difference that the integral over the post-shock volume $V_{\mathrm{ps}}$
extends from $R_{\mathrm{ns}}$ to $R_{\mathrm{s}}(\theta,\phi)$,
and that only zones where the total energy is positive are included in the integral.
Figure~\ref{fig:explosion_energies_s90_2D_3D} shows the diagnostic explosion energies
for all 2D (top panel) and 3D (bottom panel) simulations of the s9.0 model.
Our 3D simulations result in diagnostic explosion energies that almost reach (models `FMD L' and `FMD H') and even exceed (model `RbR+ L')
a value of ${4\unitspace \mathsf{x}\unitspace 10^{49} \mathrm{erg}}$, when measured at roughly ${600\unitspace\mathrm{ms}}$ after bounce
(note that explosion energies have not attained their terminal values at that time, and that model `RbR+ H' was not simulated that long).
In contrast, 2D simulations exhibit explosion energies that are systematically smaller by a factor $2$ to $3$
and start rising slightly later (in correspondence to the later outward acceleration of the shock in 2D).
A similar trend of lower explosion energies in 2D than in 3D simulations was found by \citet{2015ApJ...801L..24M},
who performed CCSN simulations of a low-mass progenitor model with a mass of ${9.6\unitspace\msun}$ and zero metallicity,
and by \citet{2015MNRAS.453..287M} for simulations of an ${11.2\unitspace\msun}$ model.

Furthermore, the diagnostic explosion energies in 2D are higher in the low-resolution compared to the high-resolution simulations
by almost ${0.8\unitspace \mathsf{x}\unitspace 10^{49} \mathrm{erg}}$,
which is a result of larger post-shock volumes in the low-resolution models after roughly ${400\unitspace\mathrm{ms}}$
(compare the evolution of angle-averaged shock radii in Figure~\ref{fig:S90_comparison_2D}).
With the small sample of 2D simulations for the s9.0 model available,
it is not clear to us whether this 2D difference is connected to resolution differences per se,
or to the non-equidistant angular grid.
In 3D, the considerably smaller differences of the diagnostic explosion energies stem from the morphology of the post-shock layer,
while the angle-averaged shock radii are very similar between simulations (see Figure~\ref{fig:radius_S90_3D}).
Model `FMD H' exhibits the most extreme shock deformation,
which is evident from the fact that the dipole mode in this model is larger than in all other models (see Figure~\ref{fig:radius_S90_3D}).
This difference is probably caused by stochastic variations,
and leads to the smallest post-shock volume and, therefore, the lowest diagnostic energy in model `FMD H'.

In summary, we find an excellent agreement of the post-bounce evolution of the s9.0 progenitor model for all of our 3D simulations.
Only small differences exist between simulations with low and high grid resolutions
with respect to the initial growth of the non-radial kinetic energies in the gain layer
and concerning the evolution of the diagnostic energies.
Differences between 3D simulations with the RbR+ approximation and the FMD neutrino-transport scheme are negligible for the s9.0 progenitor model.

\section{Conclusions} \label{sec:conclusion}

For the first time, we directly compared self-consistent CCSN simulations in 3D
with the RbR+ approximation for neutrino transport and an FMD transport scheme
to assess the influence of the two transport methods on the neutrino-heating mechanism in 3D.
Our time-dependent simulations with ``low'' and ``high'' grid resolution
(roughly ${4.5\degree}$ and ${2.25\degree}$ non-equidistant angular spacing, respectively)
for a ${9\unitspace\msun}$ and a ${20\unitspace\msun}$ progenitor model were performed
with the radiation-hydrodynamics code \textsc{Aenus-Alcar},
which implements an energy- and velocity-dependent, FMD two-moment neutrino-transport scheme
with an algebraic M1 closure relation for \nue, \nubar, and \nux,
and which is able to employ the RbR+ approximation by setting all non-radial flux components to zero.

For both progenitor models we found very good agreement of the post-bounce evolution between 3D simulations with
RbR+ and the FMD neutrino-transport scheme.
Our 3D simulations of the ${9\unitspace\msun}$ model result in successful shock revival at roughly ${300\unitspace\mathrm{ms}}$ after bounce
with diagnostic explosion energies reaching ${4\times10^{49}\unitspace\mathrm{erg}}$
(and still growing by ${\sim10^{50}\unitspace\mathrm{erg}\unitspace\mathrm{s}^{-1}}$) when evaluated at ${600\unitspace\mathrm{ms}}$ after bounce.
In contrast, with our simplified neutrino opacities and small initial seed perturbations
(used to initiate the growth of hydrodynamic instabilities),
all 3D simulations of the ${20\unitspace\msun}$ model fail to revive the stalled shock wave
until at least ${400\unitspace\mathrm{ms}}$ after bounce.

These findings contrast results from 2D simulations in which the RbR+ approximation has been shown to foster shock revival
in models that featured strong sloshing motions of SASI along the polar symmetry axis \citep{2016ApJ...831...81S,2018MNRAS.481.4786J}.
While our 2D simulations of the ${20\unitspace\msun}$ model are dominated by such axial sloshing motions,
the 3D simulations exhibit alternating phases of convective and SASI activity and,
thus, show a substantially different post-bounce evolution than corresponding 2D simulations.
In contrast, simulations of the ${9\unitspace\msun}$ model are dominated by convection in both 2D and 3D
and do not exhibit any obvious SASI activity.

Considering surface-integrated neutrino luminosities and mean energies,
we found very good agreement of the results for all neutrino species between the 3D simulations of each progenitor model.
Although angular variations of the radial neutrino fluxes differ between simulations with RbR+ and FMD,
as a detailed analysis of the ${20\unitspace\msun}$ model revealed
(confirming previous investigations by \citealt{2015ApJS..216....5S}),
the 3D simulations systematically exhibit significantly smaller variations than corresponding 2D cases.
Furthermore, the time-averaged (over typically ${10\unitspace\mathrm{ms}}$ or longer)
deviations of the radial fluxes and the neutrino-heating rates from their angular averages are in good agreement
between 3D simulations with RbR+ and FMD.
Moreover, the neutrino-heating rates near the instantaneous SASI direction agree considerably better
between 3D simulations with RbR+ and FMD than in 2D models,
in which fluctuations of the heating rates in the polar directions are much larger with RbR+ transport.
Therefore, aside from stochastic fluctuations in the shock radius,
the overall evolution for both progenitor models is very similar in 3D with FMD and RbR+ transport.
Our results show that the differences between 3D models with different resolution is larger
than the differences with the two transport treatments.

In conclusion, the post-bounce evolution of 3D simulations with the RbR+ approximation and the FMD neutrino-transport scheme
agrees much better than between corresponding 2D models.
These results back up the use of RbR+ as transport description in 3D supernova modeling.
The RbR+ approximation is beneficial regarding the parallel efficiency,
because the individual rays can be calculated almost independently with only small communication overhead between MPI tasks.
However, the preferred choice of neutrino-transport approximation strongly depends on the physical question that should be answered.
For example, the FMD scheme provides approximative representations of the non-diagonal elements of the neutrino-pressure tensor,
which may be relevant for some problems but cannot be directly extracted from a RbR+ treatment.
Another example are fast rotating progenitor models, where the use of the RbR+ approximation is disfavored,
when the proto-neutron star becomes strongly deformed and non-radial flux components cannot be neglected
in the neutrino-decoupling region near the neutron-star surface.
In the study by \citet{2018ApJ...852...28S} the rotation rate and neutron-star deformation were not sufficiently extreme
to worry about deficiencies of the RbR+ treatment.
A clear picture of shortcomings in dependence of the neutron-star rotation will require future comparisons
of rapidly rotating models with RbR+ transport versus FMD transport.
However, the good physical accuracy of the M1 FMD scheme in such situations is also not guaranteed and needs to be verified
by full Boltzmann neutrino transport.

In general, an ultimate assessment of the consequences of transport approximations entering both the FMD as well as RbR+ two-moment treatments
will require future, computationally much more demanding time-dependent 2D and 3D CCSN simulations with Boltzmann neutrino transport.
For first low-resolution steps in this direction, see \citet{2015ApJS..216....5S} and \citet{2018ApJ...854..136N}.

\acknowledgments

We are grateful to Robert Bollig for providing the low-temperature extension of the SFHo equation of state.
At Garching, this project was supported
by the European Research Council through grant ERC-AdG No.\ 341157-COCO2CASA,
and by the Deutsche Forschungsgemeinschaft
through Sonderforschungbereich SFB 1258 ``Neutrinos and Dark Matter in Astro- and Particle Physics'' (NDM)
and the Excellence Cluster Universe (EXC 153; http://www.universe-cluster.de/).
OJ acknowledges support by the Special Postdoctoral Researchers (SPDR) program and the iTHEMS cluster at RIKEN.
MO acknowledges support from the European Research Council (grants
CAMAP-259276 and EUROPIUM-667912), from the Deutsche Forschungsgemeinschaft through Sonderforschungsbereich SFB 1245
``Nuclei: From fundamental interactions to structure and stars'',
and from the Spanish Ministry of Economy and Competitivity and the Valencian Community under grants
AYA2015-66899-C2-1-P and PROMETEOII/2014-069, respectively.
Computer resources for this project have been provided by the Leibniz Supercomputing Centre (LRZ) under grant pr62za,
and by the Max Planck Computing and Data Facility (MPCDF) on the HPC system Hydra.

\software{\textsc{Aenus-Alcar} \citep{{2008ObergaulingerPhD},2015MNRAS.453.3386J,2018MNRAS.481.4786J},
NumPy and Scipy \citep{Oliphant2007},
IPython \citep{Perez2007},
Matplotlib \citep{Hunter2007}.}

\appendix

\section{Calculation of the $\chi$ Parameter}\label{sec:appendix_chi_parameter}

The $\chi$ parameter \citep{2006ApJ...652.1436F} is used to analyze the conditions
for the growth of convection in the post-shock layer.
It essentially relates the advection timescale, which can be approximated by
the ratio of the radial cell width $\mathrm{d}{r}$ over the radial velocity $v_r$,
to the timescale of convective growth,
given by the inverse of the Brunt-V\"ais\"al\"a frequency ${\omega_{\mathrm{BV}}}$,
\begin{equation}\label{eqn:chi_parameter_general}
\chi =
\int \frac{\omega_{\mathrm{BV}}}
{\left| v_r \right|} \dint{r} .
\end{equation}
The radial integral in Equation~\eqref{eqn:chi_parameter_general} extends over the gain layer, i.e.,
from the gain radius to the shock radius.
According to \citet[][assuming the linear regime for the growth of perturbations]{2006ApJ...652.1436F},
the $\chi$ parameter must exceed a critical value of $\sim 3$ for the growth of convection in the gain layer.
In formulating this criterion, \citet{2006ApJ...652.1436F} considered a spherically symmetric gain-layer configuration.
However, since in reality the gain layer has a complex multi-dimensional structure for most of the time,
non-trivial ambiguities arise as to how to compute and angle average the quantities entering Equation~\eqref{eqn:chi_parameter_general}.
As a consequence, as was already demonstrated by \citet{2014MNRAS.440.2763F},
different calculation methods may lead to significantly different values of $\chi$.
In order to assess the sensitivity of $\chi$ with respect to its practical computation,
in this appendix we systematically compare various possible methods for computing $\chi$.
All following tests are based on the 3D model `S20 FMD L' of the present study
(for a discussion of this model, see Section~\ref{sec:s20_3D}).

\begin{figure*}
    \includegraphics[width=\textwidth]{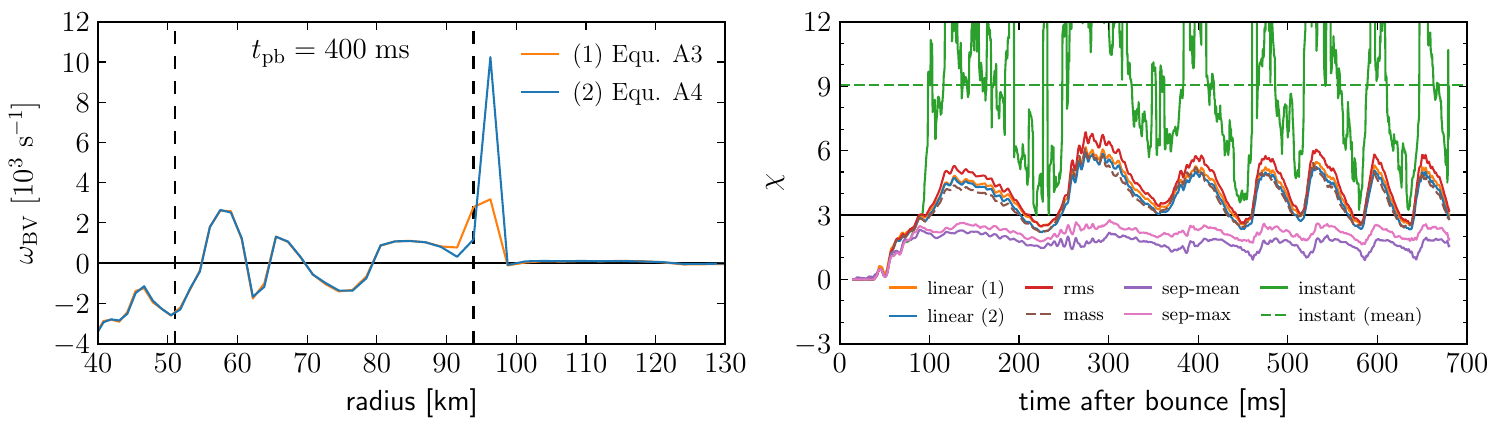}
    \caption{Radial profiles of the Brunt-V\"ais\"al\"a frequency $\omega_{\mathrm{BV}}$ (left panel),
            evaluated at arbitrarily chosen angles of $\theta = 6\unitspace\degree$ and $\phi = 2.25\unitspace\degree$
            for a snapshot of model `S20 FMD L' at a post-bounce time of $t_{\mathrm{pb}} = 400\unitspace\mathrm{ms}$,
            and the $\chi$ parameter as a function of time after bounce (right panel) for the same model.
            The line colors and line styles correspond to different methods that we use to calculate the Brunt-V\"ais\"al\"a frequency
            and the $\chi$ parameter (see labels and Table~\ref{tab:chi_parameter}).
            The horizontal lines depict the critical values for $\omega_{\mathrm{BV}} = 0$ (left panel)
            and $\chi = 3$ (right panel),
            and the dashed lines in the left panel display the gain and shock radii,
            $R_{\mathrm{g}}$ and $R_{\mathrm{s}} (\theta = 6\unitspace\degree, \phi = 2.25\unitspace\degree)$, respectively.
            The large peak in $\omega_{\mathrm{BV}}$ just in front of the shock is nonphysical (see text)
            and is not included in the radial integral for the calculation of $\chi$.
            The $\chi$ parameter was smoothed by running averages of ${5\unitspace\mathrm{ms}}$.
            \label{fig:chi_parameter}}
\end{figure*}

Before considering the $\chi$ parameter itself, we start by comparing two possibilities for the numerical evaluation
of the local Brunt-V\"ais\"al\"a frequency $\omega_{\mathrm{BV}}$.
We use a definition that is based on the criterion for Ledoux convection,
\begin{equation}\label{eqn:brunt_vaisala}
\omega_{\mathrm{BV}} = \mathrm{sign} \left( C_{\mathrm{led}} \right)
\sqrt{ \frac{\partial \phi_{\mathrm{grav}}}{\partial r} \rho^{-1}
\left| C_{\mathrm{led}} \right|} \; > 0,
\end{equation}
with the gravitational potential $\phi_{\mathrm{grav}}$ and the density $\rho$.
The Ledoux criterion is given by
\begin{equation}\label{eqn:ledoux_criterion}
C_{\mathrm{led}} =
\left( \frac{\partial \rho}{\partial s} \right)_{p, Y_{e}} \frac{\partial s}{\partial r} +
\left( \frac{\partial \rho}{\partial Y_{e}} \right)_{p, s} \frac{\partial Y_{e}}{\partial r} > 0,
\end{equation}
with entropy per baryon $s$, electron fraction $Y_{e}$, and pressure $p$.
For this definition of the Ledoux criterion convectively unstable regions fulfill
the condition $C_{\mathrm{led}} > 0$ or ${\omega_{\mathrm{BV}} > 0}$.
The thermodynamic derivatives in Equation~\eqref{eqn:ledoux_criterion} with respect to entropy and electron fraction
need to be calculated from the equation of state.
Another way to calculate the Ledoux criterion is given by
(see, e.g., \citealt{2014HuedepohlPhD}, Equation~(2.35) and Appendix B therein)
\begin{equation}\label{eqn:ledoux_criterion_den_and_pre}
C_{\mathrm{led}} =
\frac{\partial \rho}{\partial r} -
\frac{1}{c_{\mathrm{s}}^2} \frac{\partial p}{\partial r} > 0,
\end{equation}
with the local speed of sound $c_{\mathrm{s}}$.
This analytically identical alternative definition is based on radial gradients of density and pressure and, therefore,
does not need additional information from the equation of state.
Both definitions of $C_{\mathrm{led}}$ result in basically identical values for the Brunt-V\"ais\"al\"a frequency,
as can be seen in the left panel of Figure~\ref{fig:chi_parameter},
which shows radial profiles of $\omega_{\mathrm{BV}}$
for method 1 (orange line, based on Equation~\eqref{eqn:ledoux_criterion})
and for method 2 (blue line, based on Equation~\eqref{eqn:ledoux_criterion_den_and_pre}).
The large peak in $\omega_{\mathrm{BV}}$ for method 2 just in front of the shock (right dashed line) is
a physically irrelevant numerical artifact because it results from the numerical description of the shock front as a discretized flow structure.
For this reason, we do not include this region in the radial integral in Equation~\eqref{eqn:chi_parameter_general}.
The results of the above comparison, first, support the correct implementation of both computation methods in our analysis routines,
and second, confirm that numerical differences between both methods are small enough
to render both methods equivalent for the practical evaluation of $\chi$.

We now consider different computation methods for the $\chi$ parameter.
The right panel of Figure~\ref{fig:chi_parameter} shows the $\chi$ parameter as a function of the time after bounce,
obtained using the variations of calculation methods that are described in the following
(see Table~\ref{tab:chi_parameter} for an overview of the methods).
We identify three different classes of possible outcomes (corresponding to the three distinct groups of lines),
which result from different realizations regarding the angular average and the radial integral in Equation~\eqref{eqn:chi_parameter_general}.
In the first class (consisting of methods ``linear (1)'', ``linear (2)'', ``rms'', and ``mass'') $\chi$ is computed as
\begin{equation}\label{eqn:chi_parameter_class_1}
\chi =
\int \frac{\left< \omega_{\mathrm{BV}} (r, \theta, \phi) \right>}
{\left| \left< v_r (r, \theta, \phi) \right> \right|} \dint{r} ,
\end{equation}
where angle brackets denote angular averaging over angles $\theta$ and $\phi$.
Instead of averaging the three-dimensional Brunt-V\"ais\"al\"a frequency ${\omega_{\mathrm{BV}} (r, \theta, \phi)}$,
it is possible to obtain a radial profile ${\omega_{\mathrm{BV}}} (r)$ from angle-averaged quantities
$\left< \phi_{\mathrm{grav}} \right>$, $\left< \rho \right>$, $\left< c_{\mathrm{s}} \right>$, and $\left< p \right>$.
Using the radial profile ${\omega_{\mathrm{BV}}} (r)$ we obtain the second class (consisting of methods ``sep-mean'' and ``sep-max'') by
\begin{equation}\label{eqn:chi_parameter_class_2}
\chi =
\int \frac{\omega_{\mathrm{BV}} (r)}
{\left| \left< v_r (r, \theta, \phi) \right> \right|} \dint{r} .
\end{equation}
The third class (method ``instant'') is obtained by performing the angle average \emph{after} the radial integral,
\begin{equation}\label{eqn:chi_parameter_class_3}
\chi = \left<
\int \frac{\omega_{\mathrm{BV}} (r, \theta, \phi)}
{\left| v_r (r, \theta, \phi) \right|} \dint{r} \right> .
\end{equation}
In addition to these three main classes we investigated further variations of the calculation of $\chi$.
We describe these variations in great detail in the following paragraphs.
An overview of all seven methods that we used to calculate the $\chi$ parameter can be found in Table~\ref{tab:chi_parameter}.

Starting with the first class (Equation~\eqref{eqn:chi_parameter_class_1}),
we investigate the influence of the calculation method for the Ledoux criterion (see discussion above).
The excellent agreement between both calculation methods for $C_{\mathrm{led}}$ can also be seen from results for the $\chi$ parameter:
the values for method 1 (orange line; see Equation~\eqref{eqn:ledoux_criterion}) agree very well
with those obtained from method 2 (blue line; see Equation~\eqref{eqn:ledoux_criterion_den_and_pre}).
Both results for $\chi$ have been calculated with ``linear'' angular averaging of the Brunt-V\"ais\"al\"a frequency
as expressed in Equation~\eqref{eqn:chi_parameter_class_1}.
When performing the angular average over $\omega_{\mathrm{BV}} (r, \theta, \phi)$ we only take into account regions
(a) that lie in the gain layer with radii $R_\mathrm{g} < r < R_\mathrm{s} (\theta, \phi)$ and, thus,
we exclude the nonphysical shock artifact mentioned above,
and (b) that are locally unstable for convection, i.e.\ where $\omega_{\mathrm{BV}} (r, \theta, \phi) > 0$.
In contrast, the angular average of the radial velocity, $\left< v_r (r, \theta, \phi) \right>$,
is calculated from all zones with $R_\mathrm{g} < r < R_\mathrm{s}^{\mathrm{max}}$
with the maximum shock radius $R_\mathrm{s}^{\mathrm{max}}$ to capture the mean properties of the entire flow.
We apply volumetric angular averaging for $\omega_{\mathrm{BV}}$ and $v_r$,
i.e., we weight all values with the volume of the corresponding cells
and normalize the sum by dividing through the total volume.
Since possibly not all zones of a radial shell contribute to the angular average of $\omega_{\mathrm{BV}}$
due to the constraints $\omega_{\mathrm{BV}} (r, \theta, \phi) > 0$ and $R_\mathrm{g} < r < R_\mathrm{s} (\theta, \phi)$,
also the total volume that is used for the normalization of $\omega_{\mathrm{BV}}$ is calculated only from contributing zones.
After angular averaging, the radial integral in Equation~\eqref{eqn:chi_parameter_class_1} is formally carried out
from the gain radius $R_\mathrm{g}$ to the maximum shock radius $R_\mathrm{s}^{\mathrm{max}}$.

\begin{deluxetable}{ccccll}
    \tablecaption{Possible methods to calculate the $\chi$ parameter.
                  The volume that is considered for angle averaging and radial integration is given
                  as a range between the gain radius $R_\mathrm{g}$ and the
                  angle-dependent shock radius $R_\mathrm{s} (\theta, \phi)$,
                  the angle-averaged mean shock radius $R_\mathrm{s}$,
                  or the maximum shock radius $R_\mathrm{s}^\mathrm{max}$.
                  For the analysis in the main text of this paper,
                  we use the method ``linear (2)''.
                  \label{tab:chi_parameter}}
    \tablehead{
        \multicolumn{1}{c}{Method Name} &
        \multicolumn{1}{c}{Class} &
        \multicolumn{1}{c}{Angle Averaging} &
        \multicolumn{1}{c}{Weighting} &
        \multicolumn{1}{c}{Considered Volume} &
        \multicolumn{1}{c}{Method for $C_{\mathrm{led}}$}
    }
    \startdata
    linear (1)    &  Equation~\eqref{eqn:chi_parameter_class_1} &
                     linear            & volume & $R_\mathrm{g} < r < R_\mathrm{s} (\theta, \phi)$  &
                     1 - Equation~\eqref{eqn:ledoux_criterion}                          \\
    linear (2)    &  Equation~\eqref{eqn:chi_parameter_class_1} &
                     linear            & volume & $R_\mathrm{g} < r < R_\mathrm{s} (\theta, \phi)$  &
                     2 - Equation~\eqref{eqn:ledoux_criterion_den_and_pre}              \\
    rms           &  Equation~\eqref{eqn:chi_parameter_class_1} &
                     root-mean-squared & volume & $R_\mathrm{g} < r < R_\mathrm{s} (\theta, \phi)$  &
                     2 - Equation~\eqref{eqn:ledoux_criterion_den_and_pre}              \\
    mass          &  Equation~\eqref{eqn:chi_parameter_class_1} &
                     linear            & mass   & $R_\mathrm{g} < r < R_\mathrm{s} (\theta, \phi)$  &
                     2 - Equation~\eqref{eqn:ledoux_criterion_den_and_pre}              \\ \hline
    sep-mean      &  Equation~\eqref{eqn:chi_parameter_class_2} &
                     separate          & volume & $R_\mathrm{g} < r < R_\mathrm{s}$                 &
                     2 - Equation~\eqref{eqn:ledoux_criterion_den_and_pre}              \\
    sep-max       &  Equation~\eqref{eqn:chi_parameter_class_2} &
                     separate          & volume & $R_\mathrm{g} < r < R_\mathrm{s}^{\mathrm{max}}$  &
                     2 - Equation~\eqref{eqn:ledoux_criterion_den_and_pre}              \\ \hline
    instant       &  Equation~\eqref{eqn:chi_parameter_class_3} &
                     instantaneous     & volume & $R_\mathrm{g} < r < R_\mathrm{s} (\theta, \phi)$  &
                     2 - Equation~\eqref{eqn:ledoux_criterion_den_and_pre}              \\
    \enddata
\end{deluxetable}

For root-mean-squared averaging of $\omega_{\mathrm{BV}}$ (``rms'' in Table~\ref{tab:chi_parameter}) as described in \citet{2014MNRAS.440.2763F}
we replace in Equation~\eqref{eqn:chi_parameter_class_1} the linear angular average $\left< \omega_{\mathrm{BV}} \right>$
by the root-mean-squared angular average $\sqrt{\left< \omega_{\mathrm{BV}}^2 \right>}$.
As shown in the right panel of Figure~\ref{fig:chi_parameter},
root-mean-squared averaging increases the $\chi$ parameter by values of at most 1 (red line)
with respect to the ``linear'' averaging.
Even smaller differences with respect to the ``linear'' angular averaging method with volumetric weighting are found
when the weighting is performed with the mass of each zone (``mass'' in Table~\ref{tab:chi_parameter},
brown dashed line in Figure~\ref{fig:chi_parameter}).

The second class for the calculation of $\chi$ (Equation~\eqref{eqn:chi_parameter_class_2}) employs separately
angle-averaged quantities
$\left< \phi_{\mathrm{grav}} \right>$, $\left< \rho \right>$, $\left< c_{\mathrm{s}} \right>$, and $\left< p \right>$
to compute a spherically symmetric radial profile $\omega_{\mathrm{BV}} (r)$
(see Equations~\eqref{eqn:brunt_vaisala} and \eqref{eqn:ledoux_criterion_den_and_pre}).
Here, the angular averages are again computed with volumetric weighting and include all zones with
$R_\mathrm{g} < r < R_\mathrm{s}$ or $R_\mathrm{g} < r < R_\mathrm{s}^{\mathrm{max}}$,
depending on the method:
We compare values for $\chi$ where the radial integral (including only regions where $\omega_{\mathrm{BV}} (r) > 0$)
was performed up to the mean shock radius $R_\mathrm{s}$
(``sep-mean'' in Table~\ref{tab:chi_parameter})
and the maximum shock radius $R_\mathrm{s}^{\mathrm{max}}$
(``sep-max'' in Table~\ref{tab:chi_parameter}), respectively.
Separately averaging the quantities leads to significantly reduced values of $\chi$ not exceeding 2
(for ``sep-mean'', purple line in Figure~\ref{fig:chi_parameter})
or 3 (for ``sep-max'', pink line in Figure~\ref{fig:chi_parameter}).
Since both \citet{2016ApJ...825....6S} and \citet{2018MNRAS.481.4786J} use similar methods as the ``sep-mean'' one,
this could explain the rather small values that both studies find for the $\chi$ parameter
in their 2D simulations for the same s20 progenitor.
The values of the $\chi$ parameter in this case are slightly below the critical level suggested by \citet{2006ApJ...652.1436F}.

Last, with the third class (Equation~\eqref{eqn:chi_parameter_class_3})
we consider an ``instantaneously'' averaged $\chi$ parameter (similarly to \citealt{2014MNRAS.440.2763F}),
which is obtained by angle averaging the $\chi$ parameter \emph{after} performing the radial integral
(``instant'' in Table~\ref{tab:chi_parameter}).
In this case, only zones with $\omega_{\mathrm{BV}} (r, \theta, \phi) > 0$ and $R_\mathrm{g} < r < R_\mathrm{s} (\theta, \phi)$
are considered for the radial integral.
We find for this method significantly larger values for the $\chi$ parameter (green line) than for all other methods,
with a temporal average as large as 9.1 (green dashed line).
This result roughly agrees with the findings by \citet{2014MNRAS.440.2763F},
who argue that the instantaneously averaged $\chi$ parameter does not capture the properties of the mean flow and,
thus, is not suitable to analyze the conditions for convective growth in the gain layer.

For the analysis in the present paper we use the ``linear (2)'' method,
which matches best our observations regarding the phases of growth of convection
($\chi \gtrsim 3$ in agreement with the critical value from the linear analysis by \citealt{2006ApJ...652.1436F})
and of SASI growth (small values of the $\chi$ parameter, roughly $\chi \lesssim 4$).

\bibliographystyle{abbrvnat}

\end{document}